\documentclass[aps,prb,twocolumn,amsmath,amssymb,nofootinbib,superscriptaddress,floatfix]{revtex4-1}
\usepackage{amssymb}
\usepackage{mathrsfs}

\usepackage{color}
\usepackage{graphicx}
\usepackage{subfigure}
\usepackage{srcltx}
\usepackage{mathrsfs}
\usepackage{multirow}
\usepackage{ulem}
\usepackage{mathtools}
\usepackage{umoline}
\usepackage{lipsum}
\usepackage{cleveref}

\newcommand{\verteq}{\rotatebox{90}{$\,=$}}
\newcommand{\equalto}[2]{\underset{\scriptstyle\overset{\mkern4mu\verteq}{#2}}{#1}}
\newcommand{\vertoplus}{\rotatebox{90}{$\,\oplus$}}
\newcommand{\oplusto}[2]{\underset{\scriptstyle\overset{\mkern4mu\vertoplus}{#2}}{#1}}

\def\beq{\begin{equation}}
\def\eeq{\end{equation}}

\begin{document}

\title{
Topological Phases Protected By Reflection Symmetry and Cross-cap States 
}

\author{Gil Young Cho}
\affiliation{
Institute for Condensed Matter Theory and
Department of Physics, University of Illinois at Urbana-Champaign, 1110 West Green St, Urbana IL 61801}
\affiliation{Department of Physics, Korea Advanced Institute of Science and Technology, Daejeon 305-701, Korea}

\author{Chang-Tse Hsieh}
\affiliation{
Institute for Condensed Matter Theory and
Department of Physics, University of Illinois at Urbana-Champaign, 1110 West Green St, Urbana IL 61801}

\author{Takahiro Morimoto}
\affiliation{Condensed Matter Theory Laboratory, RIKEN, Wako, Saitama, 351-0198, Japan}

\author{Shinsei Ryu}
\affiliation{
Institute for Condensed Matter Theory and
Department of Physics, University of Illinois at Urbana-Champaign, 1110 West Green St, Urbana IL 61801}

\date{\today}

\begin{abstract}
Twisting symmetries provides an efficient method to diagnose symmetry-protected topological (SPT) phases. 
In this paper, edge theories of (2+1)-dimensional topological phases protected by reflection as well as other symmetries  
are studied by twisting reflection symmetry,
which effectively puts the edge theories on an unoriented  spacetime, such as the Klein bottle.
A key technical step taken in this paper is the use of the so-called cross-cap states, 
which 
encode entirely the unoriented nature of spacetime, 
and can be obtained by rearranging the spacetime geometry and exchanging the role of space and time coordinates. 
When the system is in a non-trivial SPT phase, 
we find that the corresponding cross-cap state is non-invariant under the action of the symmetries 
of the SPT phase, but acquires an anomalous phase. 
This anomalous phase,  
with a proper definition of a reference state, on which symmetry acts trivially, 
reproduces the known classification of (2+1)-dimensional bosonic and
fermionic SPT phases protected by reflection symmetry,
including in particular the $\mathbb{Z}_8$ classification of topological crystalline superconductors 
protected by reflection and time-reversal symmetries.
\end{abstract}

\pacs{72.10.-d,73.21.-b,73.50.Fq}

\maketitle

\section{Introduction}

Symmetry-protected topological (SPT) phases are gapped phases of matter 
which are not adiabatically deformable, under a given set of symmetry conditions, 
to a topologically trivial phase.
While SPT phases do not have an intrinsic topological order 
(i.e., do not support (deconfined) fractional excitations), 
they are sharply (topologically) distinct from topologically trivial states,
such as an atomic insulator,  
which respect the same set of symmetries. 
In other words, 
the distinction between SPT and trivial phases cannot be made within Landau's theory;
SPT phases are beyond the classification of phases of matter based on broken symmetries. 
A flurry of recent theoretical works includes, 
among others,
the breakdown (or collapse) of the non-interacting classifications of fermionic SPT phases upon inclusion of interactions,  
non-trivial bosonic SPT phases and the possibility of symmetry-respecting surface topological order,
classification of interacting electronic topological insulators in three dimensions,  
and 
proposals for a possible complete classification of SPT phases.  
(For a partial and incomplete list of recent works on SPT phases, see Refs.
\onlinecite{
KaneRev, QiRev, Classification2, Kitaev2009, Turner2013, SenthilReview,
Fidkowski2010, Fidkowski2011,
Ryu2012, Qi2012, yao2013interaction,  
CWang2014, Fidkowski2013, Metlitski2014, 
Vishwanath2013, Geraedts2014, Lu2012a, Cho2014, Cappelli2013,
Chen2013, Kapustin2014a, Kapustin2014b, Kapustin2014c}.)

One of the most efficient and powerful methods to study SPT phases
is to \textit{twist} or \textit{gauge} symmetries protecting SPT phases. 
\cite{Levin2012, Ryu2012, Sule2013}  
Quite generically, global symmetries in quantum field theories can be twisted,
i.e., can be used to define twisted boundary conditions.
It was proposed that the twisted theory can be used to diagnose the original SPT phases,
i.e., to judge whether or not the original theory is symmetry-protected, and distinct from topologically trivial phases. 
More specifically, 
once twisted, the edge theory of an SPT phase
suffers from various kinds of \textit{quantum anomalies}, 
such as a global anomaly under large $U(1)$ gauge transformations, 
or a global gravitational anomaly.
\cite{Ryu2012, Sule2013, Hsieh2014}
On the other hand, 
gauging (non-spatial) symmetries effectively deconfines a set of quasiparticles (anyons).
The fractional statistics of the braiding in the gauged theory can be used to diagnose 
the original SPT phases. 
\cite{Levin2012, Lu2012b}

Twisting non-spatial unitary symmetries of SPT phases is by now reasonably well-understood. 
Toward further developments of methodologies to SPT phases, 
it is necessary to extend the twisting or gauging procedure to spatial and/or antiunitary symmetries, 
such as parity symmetry ($P$-symmetry) and time-reversal ($T$-symmetry). 
\cite{Hsieh2014, Kapustin2014a,Kapustin2014b, Kapustin2014c, Chen_Gauging_Tsymmetry}
The purpose of this paper is to provide an efficient and intuitive method 
to diagnose SPT phases protected by a spatial symmetry such as parity (reflection). 
We will study (1+1)-dimensional [(1+1)d] non-chiral gapless theories with spatial symmetries, 
which are the edge theories of the corresponding (2+1)d bulk SPT phases protected by 
parity ($P$) symmetry or parity combined with a unitary non-spatial symmetry 
such as $CP$-symmetry (parity symmetry combined with charge conjugation). 
In typical situations, 
in addition to $P$- or $CP$-symmetry, 
there are other non-spatial symmetries 
such as $U(1)$ symmetry or time-reversal ($T$) symmetry.

Our starting point is Ref. \onlinecite{Hsieh2014}, 
where twisting parity within edge theories of SPT phases (protected by parity) has been used to diagnose the edge theories. 
This twisting procedure leads to theories that are effectively defined on an unoriented  manifold, e.g., the Klein bottle,
and thus suggests an interesting link
between SPT phases and so-called “orientifold field theories” - type of theories discussed
in unoriented superstring theory.
\cite{Hori2003, Blumenhagen2009,Callan1987,Sagnotti1988, PolchinskiCai1988, Horava89,Angelantonj02,Brunner02, Brunner03, Yamaguchi} 
We make one further step in this paper 
by making use of the so-called \textit{cross-cap states} 
in formulating orientifold field theories. 
Cross-cap states are quantum states in field theories (conformal field theories) obtained by twisting parity symmetry
and encode, quantum mechanically, the unoriented topology; 
the fact that the theory is put on an unoriented  surface is entirely encoded in the cross-cap states. 
They are akin to boundary states in boundary conformal field theories, which are obtained by exchanging 
the role of space and time coordinates. 
One promising aspect of cross-cap states is that they
are formulated and constructed fully in terms of many-body physics, and hence are expected to
capture the effects of interactions.

We will identify quantum anomalies in the edge theories of SPT phases
as the non-invariance of a cross-cap state 
under the action of the other symmetry than parity.
Correspondingly, a bulk theory which supports an anomalous edge theory  
is diagnosed as a non-trivial SPT phase. 
Our procedure to diagnose an edge theory with parity symmetry  can be summarized as follows:
\begin{itemize}
 \item[(i):] 
The edge theory of a given bulk theory is put on an unoriented  spacetime, the Klein bottle.
In practice, this can be achieved by twisting boundary conditions (orientifold procedure).

\item[(ii):] 
The edge theory on the Klein bottle is then quantized.
The cross-cap state corresponding to the twisting is constructed. 

\item[(iii):] 
In the quantized theory, 
the effects of other symmetry, such as time-reversal, fermion number parity, etc. 
are studied.
When these symmetries are anomalous, i.e., when the cross-cap state 
is non-invariant under these symmetries, 
the edge theory cannot be gapped while preserving the symmetries. 
The corresponding bulk phase is symmetry-protected and
topologically distinct from a trivial phase. 
\end{itemize}

In the previous work,
\cite{Hsieh2014} 
the edge theories of SPT phases with 
$P \rtimes U(1)_A$ symmetry and those with $CP \rtimes U(1)_V$ symmetry have been studied. 
The partition function on the Klein bottle was shown to be anomalous (non-invariant) 
upon threading a unit flux of the $U(1)$ symmetry. 
In fact, the partition function acquires the anomalous $(-1)$-sign,
implying the ${\mathbb Z}_{2}$ classification of these SPT phases. 
The proposed formulation in this paper in terms of cross-cap states
reproduces these results for SPT phase with $P\rtimes U(1)_A$ and $CP\rtimes U(1)_V$.  
In addition, 
this formulation in terms of cross-cap states offers 
a few technical advantages.
First, 
the new formulation allows 
us to discuss SPT phases protected by a broader set of symmetries than
$P\rtimes U(1)_A$ or $CP\rtimes U(1)_V$.
That is, we do not need a continuous $U(1)$ symmetry and a large $U(1)$ gauge transformation. 
Second, it is not necessary to compute the full (symmetry-twisted) partition function; all information
necessary to diagnose edge theories are encoded in cross-cap states.

The rest of the paper is organized as follows. 
In Sec.\ \ref{Crosscap states and SPT phases},
some generalities on twisting/gauging parity symmetries and finding cross-cap states are presented. 
We further demonstrate that non-invariance of the cross-cap states under symmetry operations
is related to the non-invariance of the partition function, i.e., the quantum anomalies. 
We also draw some parallel between 
twisting non-spatial symmetries (so-called orbifold procedure)
and twisting parity symmetries (orientifold).
In Sec.\  \ref{Bosonic SPT phases},
we apply our strategy to the bosonic SPT phases protected by $P$- or $CP$-symmetry together with other symmetries,
and find that the non-invariance of the cross-cap states reproduces 
all the non-trivial SPT phases found in Ref. \onlinecite{Hsieh_CPT}. 
In Sec.\  \ref{Real fermionic SPTs}, 
we discuss real fermionic SPT phases (topological superconductors) 
protected by parity symmetry. 
By identifying quantum anomalies (anomalous phases) in the cross-cap states,
the results consistent with 
known microscopic analysis of gapping potentials are obtained.

\section{Crosscap states and SPT phases}
\label{Crosscap states and SPT phases}

In this section, some generalities of our approach to SPT phases are presented. 
We start by briefly reviewing twisting and gauging non-spatial unitary symmetries of SPT phases,
in particular within (1+1)d edge theories of (2+1)d SPT phases.
When the edge theories are realized at the boundary of non-trivial SPT phases,
this twisting procedure (orbifolding) reveals a conflict between different symmetries at the quantum level, 
even when they are mutually consistent at the classical level. 
This is an example of quantum anomalies, signaling the impossibility of realizing the edge theory on its own
once symmetry conditions are imposed. 
Hence, it is also indicative of the presence of non-trivial bulk states. 
Twisting unitary spatial symmetry, such as parity and $CP$ symmetry, can be discussed 
in a similar way, resulting in an unoriented  spacetime manifold.
Cross-cap states, which are quantum states fully encoding the unoriented nature of the theory, 
are introduced.
A potential conflict of other symmetries with parity/$CP$ symmetry can be studied in the 
language of cross-cap states.

\subsection{Edge theories of SPT phases}

By definition, when 
going from a SPT phase to a trivial phase 
in a phase diagram by changing parameters in the system's Hamiltonian, 
one inevitably encounters a quantum phase transition, if the symmetry conditions are strictly enforced. 
This in turn implies that if an SPT phase is spatially proximate to a trivial phase, there should be a gapless
state localized at the boundary between the two phases; this critical state can be thought of as
a “phase transition” occurring locally in space, instead of the parameter space of the Hamiltonian. 
As implied by this construction, the edge state of a non-trivial SPT phase
should never be removable (completely gapped) if the symmetries are strictly imposed. 
(It should be noted, however, that there are other interesting possibilities for symmetric 
surface states for (3+1)d SPT phases. 
\cite{Vishwanath2013, xuludwig, xu3dspt,Geraedts2014}) 
Hence, this critical boundary state signals the topological distinction between the SPT and trivial phases,
and many properties of SPT phases can be extracted from their boundary physics. 
For example, by inspecting under which symmetry conditions a given edge theory is stable/unstable, one can
predict under which symmetry conditions a given phase can be a SPT phase.

Although studying the boundary instead of the bulk reduces the dimensionality of the problem,
it is still not straightforward to judge if a given state is topological or not. 
In principle, one could enumerate all possible symmetry-allowed perturbations within the edge theory,  
which can potentially gap out the edge. 
Without any guiding principle, however, such “brute force” approach is quite cumbersome, 
and also, more fundamentally, does not provide any intuition on the physics of SPT phases. 
Hence it is necessary to have an efficient and illuminating
guiding principle for diagnosing topological properties of phases of matter with symmetries.

\subsection{Symmetry twist and conflicting symmetries}

Our approach to diagnose non-trivial SPT phases is to identify quantum anomalies within edge theories. 
We illustrate our strategy by considering a fermionic SPT phase protected by 
$U_c(1) \times U_s (1)$ symmetry, where $U_{c/s}(1)$ refers to $U(1)$ symmetry associated with the conservation 
of electromagnetic charge and $z$-component of spin $S_z$, respectively.  
This phase is akin to (2+1)d time-reversal symmetric topological insulators (the quantum spin Hall effect), 
although because of the imposed conservation of $S_z$, 
states with  $U_c(1)\times U_s(1)$ symmetry are classified by two integral topological invariants
(i.e., ``charge'' and ``spin'' Chern numbers). 
We will focus on the case of vanishing charge Chern number,
and hence the allowed phases are classified by the integer-valued spin Chern number.

Consider the following Hamiltonian describing the edge of the $S_z$-conserving 
quantum spin Hall phase with $U_c(1) \times U_s(1)$ symmetry defined on a circle of circumference $\ell$
\begin{align}
H = \int^{\ell}_0 dx\, 
\left[
\psi^{\dagger}_{\uparrow} (- v i\partial_{x})\psi_{\uparrow} + \psi^{\dagger}_{\downarrow} (+ vi \partial_{x})\psi_{\downarrow}
\right],  
\label{intro3} 
\end{align}
where 
$x\in [0, \ell]$ is the spatial coordinate of the edge, 
$\psi_{\uparrow/\downarrow}$ represents the fermion creation operator for up/down spin,
and 
$v$ is the Fermi velocity. 
Any global symmetry in quantum field theories can be twisted
(i.e., can be used to twist boundary conditions).
We choose $U_c(1)$ symmetry to twist the boundary condition as  
\begin{align}
\psi_{s}(x) = e^{2\pi i\alpha}
\psi_{s}(x+\ell),
\quad 
s \in \{ \uparrow, \downarrow \}. 
\label{charge bc}
\end{align}
Let the ground state with the twisted boundary condition be $ |\mathrm{GS} \rangle_{\alpha}$,
which satisfies
\begin{align}
\left[ \psi_{s}(x) - e^{2\pi i\alpha}\psi_{s}(x+\ell) \right] |\mathrm{GS} \rangle_{\alpha} = 0, 
\quad s \in \{ \uparrow, \downarrow \}. 
\end{align}

Observe that the boundary condition (\ref{charge bc}) is invariant under global $U(1)_c$ transformations
\begin{align}
0&=
\psi_{s}(x) - e^{2\pi i\alpha} \psi_{s}(x+\ell)
\nonumber \\
&=
e^{i \phi Q} \left[ \psi_{s}(x) - e^{2\pi i \alpha} \psi_{s}(x+\ell) \right]e^{-i\phi Q}, 
\end{align}
where $Q$ is the total charge operator associated to $U(1)_c$ symmetry,
and $\phi$ is an arbitrary real parameter. 
Similarly, 
the boundary condition (\ref{charge bc}) is invariant under the global $U(1)_s$ 
transformation generated by $e^{i \phi S_z}$
where $S_z$ is the total ``charge'' associated to $U(1)_s$ symmetry. 
A crucial observation now is that,
while 
the boundary condition (\ref{charge bc}) is invariant under the global $U(1)_s$,  
the corresponding ground state may carry an ``anomalous'' $S_z$ quantum number, 
\begin{align}
e^{i Q \phi} |\mathrm{GS} \rangle_{\alpha} & = e^{i \times 0 \times \phi } |\mathrm{GS} \rangle_{\alpha} = |\mathrm{GS} \rangle_{\alpha} ,
\nonumber\\ 
e^{i S_z \phi }|\mathrm{GS} \rangle_{\alpha} & = e^{i\times 2\alpha \times \phi} |\mathrm{GS} \rangle_{\alpha} = 
e^{2i\alpha \phi} |\mathrm{GS} \rangle_{\alpha}. 
\end{align}
(The quantum numbers of the ground state can be computed explicitly, by using, for example, bosonization and identifying
an operator corresponding to the ground state $|\mathrm{GS}\rangle_{\alpha}$. 
More precisely, the operator is given by $\sim \exp(i \alpha (\varphi_L + \varphi_R))$,
where the left- and right-moving electrons are identified as $\psi_{L/R} \sim \exp(\pm i\varphi_{L/R})$. 
For details, see Ref.\ \onlinecite{Cho2014}.) 
The anomalous phase is nothing but chiral anomaly;
in the presence of a $U(1)_c$ magnetic flux  twisting the boundary condition, 
the $S_z$ quantum number, which is conserved at the classical level,
is not conserved at the quantum level.  
Hence, only way to reconcile the $U(1)_c\times U(1)_s$ symmetry and quantum mechanics is 
to realize this (1+1)d theory as a boundary theory of a higher-dimensional system, 
a (2+1)d bulk SPT phase respecting the $U(1)_c \times U(1)_s$ symmetry. 

Let us now be more general.
Consider an edge theory, which is written in terms of a fundamental quantum field $\Phi(x)$
and is defined on a spatial circle $x \sim x+\ell$.
Suppose there is a set of non-spatial unitary symmetry operators, ${G}$,
which leave the edge theory invariant. 
We consider a twisting boundary condition by a group element ${g}_1 \in {G}$, 
\begin{align}
 \Phi(x+\ell) = \mathscr{G}^{\ }_1 \Phi(x) \mathscr{G}^{-1}_1 =  U_{g_1}\cdot \Phi (x), 
 \label{twisting g1}
\end{align}
where $\mathscr{G}_1$ is the operator which implements a symmetry operation $g_1$ in
the Hilbert space of the edge theory, and 
$U_{g_1}$ is a unitary matrix acting on the internal index of the field $\Phi(x)$, 
i.e., a unitary matrix representation of the symmetry.
(The field $\Phi(x)$ can carry a set of indices representing internal degrees of freedom, which are suppressed in
the equation above and in the following.) 
With twisting, 
states in the Hilbert space, 
the ground state $|\mathrm{GS}\rangle_{g_1}$ in particular, 
obey 
\begin{align}
\left[
 \Phi(x + \ell) - U_{g_1} \Phi(x)
\right]
|\mathrm{GS} \rangle_{g_1} =0. 
\end{align}
As a next step, we consider the action of another symmetry $g_2 \in G$.
($g_2$ can be equal to $g_1$, which is the situation relevant to the quantum Hall effect,
but in our examples below, $g_2\neq g_1$.)
At the classical level, the $g_1$-twisted boundary condition (\ref{twisting g1})
may be invariant under $g_2 \in {G}$,
\begin{align}
0&=
\Phi(x + \ell) - U_{g_1} \Phi(x)
\nonumber \\
&=
\mathscr{G}^{\ }_2 \left[ \Phi(x + \ell) - U_{{g}_1} \Phi(x) \right] \mathscr{G}^{-1}_2.
\end{align}
When this is the case,
one may expect the twisted theory after quantization is invariant under $g_2$ as well.
In particular, one expects the partition function and/or the ground state of the twisted edge theory 
is invariant under $g_2$. 
When this expectation is betrayed, there is a quantum anomaly.

Before leaving this subsection, 
we comment on a connection between
the twisting procedure and orbifolding/gauging.
Gauging and orbifolding 
have a similar (identical) effect in that we focus on a gauge singlet ($G$-invariant) sector of the theory 
[although the gauging means in general imposing the singlet condition locally (e.g., at each site  of a lattice), while
the projection in orbifolding is enforced only globally].  
Let us again consider an edge theory with symmetry group $G$.  
By state-operator correspondence in quantum field theories,
corresponding to the state $|\mathrm{GS}\rangle_g$ in the twisted theory,
there is an operator, the twist operator $\sigma_{g}$,
which, when acting on the ground state of the untwisted theory, 
creates $|\mathrm{GS}\rangle_g$ 
\cite{CFTbook, PolchinskiStringbook,Ginsparg91}:
\begin{align}
|\mathrm{GS}\rangle_g= \sigma_{g} (0)|0\rangle. 
\end{align}
(The location of the insertion of the operator $\sigma_{g}$ is taken to be the origin in the radial quantization.) 
The twist operator, when inserted in correlation functions, 
implements the symmetry twist by $g$, 
and satisfy the following algebraic relation with $\Phi(x)$
on the complex plane:
\begin{align}
 \Phi(z,\bar{z}) \sigma_g(w, \bar{w}) 
 =
 \sigma_g(w,\bar{w}) U_g \cdot \Phi(z,\bar{z}).  
\end{align}

Starting from the original, untwisted, theory,
one can now consider including the ground states with twisted boundary conditions
to define an extended  theory. 
In the extended theory, 
the ground states with twisted boundary conditions 
(the ground states in the ``twisted sectors''),
and hence the corresponding twist operators, 
are considered as an excitation.
This procedure to generate a new theory from the untwisted theory is called orbifold. 

Now, by further invoking bulk-boundary correspondence, 
there is a corresponding bulk excitation (anyon). 
Bulk statistical properties of the gauged theory can be read off 
from the operator product expansions and fusion rules obeyed by
the twist operator(s).
Now a given symmetry group $G$ can be implemented in various different ways
in different SPT (and trivial) phases protected by $G$
leading to different choices of $U_g$, and to different twist operators. 
By studying statistical (braiding) properties of the twist operator(s),
one can distinguish different ungauged (original) theories.  
\cite{Levin2012}

\subsection{Spatial symmetry}

When considering SPT phases protected by spatial symmetries, 
one can consider twisting the spatial symmetries.  
Let us consider a parity symmetry:
\begin{align}
\mathscr{P}\Phi(x)\mathscr{P}^{-1}
 =  U_{{P}} \Phi(\ell-x), 
\end{align}
where the space is defined on a circle $x \sim x+ \ell$. 
Twisting by parity symmetry can be introduced in the following way. 
\cite{Brunner02, Brunner03, Yamaguchi,Horava89, Hori2003, Blumenhagen2009}

\paragraph{Loop channel}

We consider Euclidean spacetime $[0, \ell] \times [0, \beta]$  parameterized by $(x_1,x_2)$,
where $x_1$ is the spatial coordinate and $x_2$ is the imaginary time coordinate. 
In the path integral picture, the fields obey the following twisted boundary conditions 
\begin{align}
 \Phi(x_1, x_2+ \beta) 
 &=
 \mathscr{P} \Phi(x_1, x_2)\mathscr{P}^{-1}
 =
 U_{P} \Phi(\ell-x_1, x_2),
 \nonumber \\%%%%%
 \Phi(\ell, x_2) &= \mathscr{G} \Phi(0,x_2) \mathscr{G}^{-1}, 
\label{BC_xCap}
\end{align}
in which $\mathscr{G}$ may be implementing a non-spatial unitary symmetry $g \in G$,
e.g., the fermion number parity. 
%This means that the theory projected by the parity is defined on the Klein bottle.
\cite{Hsieh2014}
In the operator language, 
twisting by parity symmetry can be introduced as a projection operation. 
This amounts to inserting the parity operator into the partition function,  
\begin{align}
 Z^{K}
 &=
 \mathrm{Tr}_{g} 
 \left[
 \mathscr{P} e^{-\beta H^{\ }_{\mathrm{loop}}(\ell) } 
 \right],
% \nonumber \\
% &= \langle C_1 | e^{- \frac{L}{2} H_c(2\beta) } |C_2\rangle 
\end{align}
where $H_{\mathrm{loop}}\equiv H$ is the Hamiltonian that generates time-translation
in the $x_2$-direction.
The trace is taken in the Hilbert space defined for the quantum field obeying the 
boundary condition $\Phi(x_1+\ell) = \mathscr{G} \Phi(x_1) \mathscr{G}^{-1}$,
as indicated by the subscript $g$.
(For later use, we also introduce another parity operator,
$\mathscr{P}'$, such that 
$\mathscr{P}^{\prime}\cdot \mathscr{P}^{-1} =\mathscr{G}$.)
This representation of the partition function with the choice of $x_2$ 
as a direction of time-evolution will be called the ''loop channel'' picture.

\paragraph{Tree channel}

As we have seen in the case of non-spatial symmetries, 
it is useful to discuss the twisted partition function 
in terms of the twist operator and the corresponding state. 
One may want to develop a similar alternative picture for the case of twisting by parity.
To this end, one needs to find a convenient and proper 
``time-slice'' that allows us to define a quantum state 
($|\mathrm{GS}\rangle_g$ in the notation of the previous section for the non-spatial symmetry),  
which implements the twisting condition.

\begin{figure}
\begin{center}
\includegraphics[width=\columnwidth]{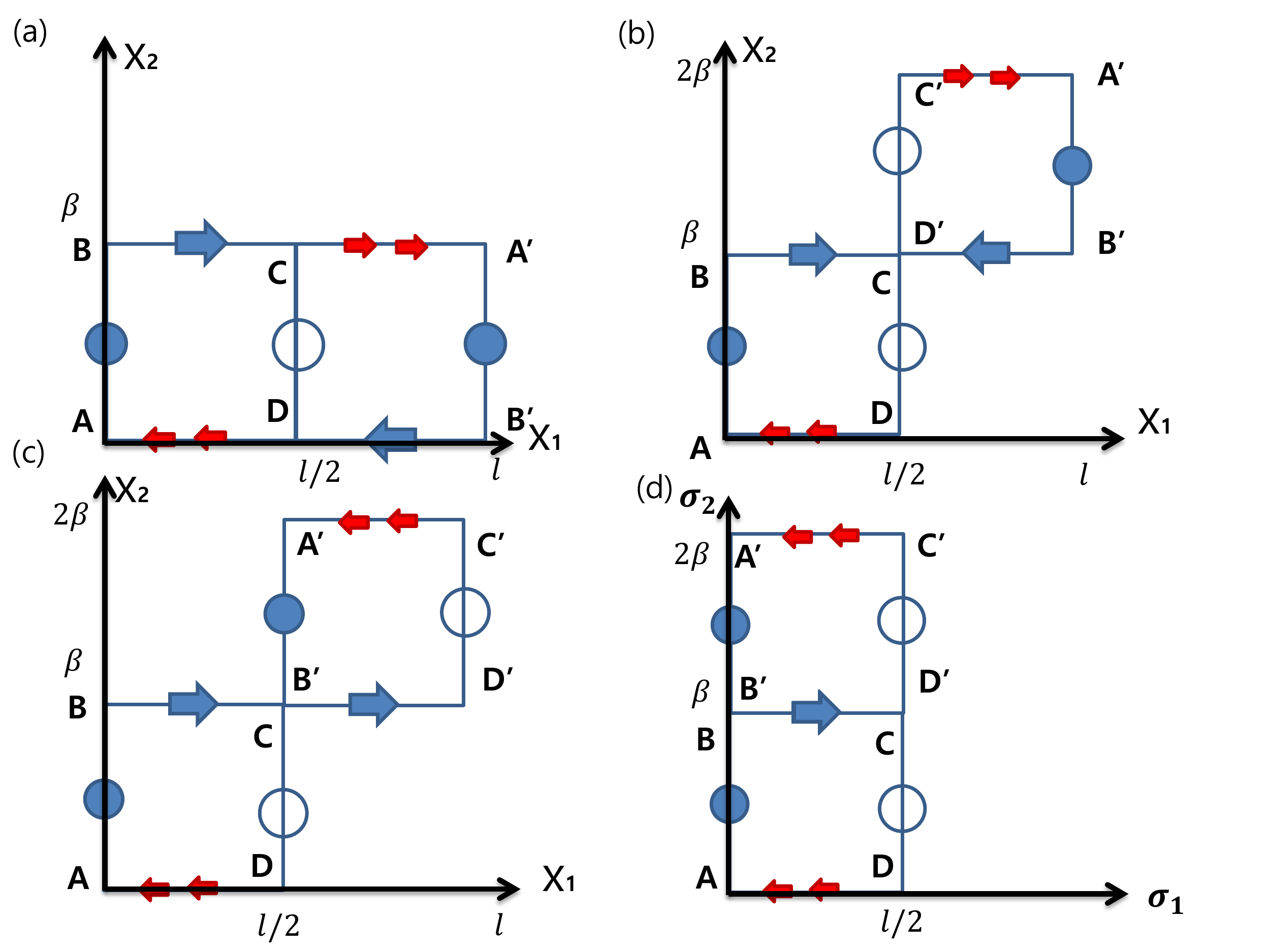}
\caption{
Rearrangement of spacetime. 
(a) The original loop channel (Euclidean) spacetime. 
Here the points $(A,B)$ and $(A',B')$ in the spacetime are identified due to the boundary conditions Eq.\ \eqref{BC_xCap}. 
The line segments with the same symbols are also identified by the boundary conditions Eq.\ \eqref{BC_xCap}. 
(b) We shift the box of the spacetime formed by $A'B'D'C'$ along $x_2$ by $\beta$. 
(c) After shifting the box, we flip the orientation of the box with respect to $x_1 = 3\ell/4$. 
(d) We slide the box along $x_1$ by $\ell/2$. 
Then we obtain the crosscap geometry. 
Notice that any $x_2 \in [0, \beta]$ is identified with its antipodal point at the slice of ``time" at $x_1 = 0$ 
and $x_1 = \ell/2$. 
By renaming the variables $x_1 \to \sigma_1$ and $x_2 \to \sigma_2$, we obtain the spacetime of the tree channel. Here the $\sigma_1$ direction is taken to be a direction of 
(fictitious) time-evolution. At the boundary of spacetime located at $\sigma_1=0$ and $\sigma_1=\ell/2$, 
there are cross-caps. 
 \label{fig:xCap}
}
\end{center}
\end{figure}

It is convenient to recall that 
any 2d compact unoriented surface without boundary can be generated from a sphere $S^2$
by adding handles and cross-caps, which can be thought of as a real projective plane $\mathbb{RP}_2$. 
In particular, 
the Klein bottle can be generated by first cutting out two discs from $S^2$,
and then identifying antipodal points of the resulting holes.   
In fact, one can rearrange the path integral on the Klein bottle 
such that the fields are now defined on $[0, \ell/2] \times [0, 2\beta]$. 
The precise steps for rearranging the spacetime is described in Fig.\ref{fig:xCap}.

We use $(\sigma_1,\sigma_2)$ to represent this rearranged spacetime. 
The fields now obey the cross-cap boundary conditions 
\begin{align}
\Phi(0, \sigma_2+\beta)
&=
U_{P'} \Phi(0, \sigma_2),
\nonumber \\%%%%%
\Phi(\ell/2, \sigma_2+\beta) 
&=
U_{{P}}  \Phi(\ell/2, \sigma_2),
\nonumber \\%%%%%
\Phi(\sigma_1, \sigma_2+2\beta)
&=
\mathscr{G}' \Phi(\sigma_1, \sigma_2)\mathscr{G}^{\prime -1}, 
\end{align}
where 
\begin{align}
\mathscr{P}^2 = \mathscr{P}^{\prime 2} =\mathscr{G}^{\prime}. 
\label{cross cap formula 1}
\end{align}
By further rotating $(\sigma_1,\sigma_2)$ coordinates by $90^{\circ}$, 
we can take $\sigma_1$ as a time-coordinate. 
In this coordinate system, the time-evolution is generated by 
a (fictitious) Hamiltonian (denoted by $H_{\mathrm{tree}}$ in the following).
%which is an ordinary one defined for an oriented system. 
%On the other hand, 
The fact that the system is defined on an unoriented  surface
is now encoded in boundary conditions at $\sigma_1=0$ and $\sigma_1=\ell/2$. 
We then introduce cross-cap states which obey these boundary conditions as:
\begin{align}
&
\left[
\Phi(0, \sigma_2+\beta)
-
U_{P'} \Phi(0, \sigma_2)
\right]|C_{P} \rangle
=0, 
\nonumber \\%%%%%
&
\left[
\Phi(\ell/2, \sigma_2+\beta) 
-
U_{{P}}  \Phi(\ell/2, \sigma_2)
\right]|C_{P^{\prime}}\rangle=0. 
\end{align}

The partition function can be written in terms of the cross-cap states as 
\begin{align}
Z^{K}=
 \langle {C}_{P^{\prime}} | e^{-\frac{\ell}{2} H^{\ }_{\mathrm{tree}} (2\beta)} 
 |{C}_{P} \rangle,  
\end{align}
where $H^{\ }_{\mathrm{tree}}$ is the Hamiltonian that generates time-translation 
in the tree channel picture.
The loop channel-tree channel duality (also known as open-closed duality)
asserts that the partition functions computed in the loop and tree channels agree.

\paragraph{Interplay with non-spatial symmetries}

By constructing the cross-cap states, 
we have ``gauged'' parity symmetry. 
As in the case of the SPT phases with non-spatial symmetries, 
we now consider the effects of another non-spatial symmetry, 
represented by $ g\in {G}$, on the cross-cap states. 
We act with $g$ on the cross-cap boundary conditions, 
\begin{align}
 &\quad 
 \left[
 \Phi(0,  \sigma_2 + \beta) 
- 
U_{{P}} \Phi(0, \sigma_2) 
\right]
| {C}_{{P}}\rangle 
=0
\nonumber \\%%%%%
 &\Rightarrow 
 \mathscr{G}
 \left[
 \Phi(0, \sigma_2 +\beta) 
- 
U_{{P}} \Phi(0, \sigma_2) 
\right]
\mathscr{G}^{-1}
\mathscr{G}
| {C}_{{P}}\rangle 
=0
\nonumber \\%%%%%
 &\Rightarrow 
 \left[
  U_{{g}} \Phi(0, \sigma_2 + \beta) 
- 
U_{{P}} U_{{g}} \Phi(0, \sigma_2) 
\right]
\mathscr{G}
| {C}_{{P}}\rangle 
=0
\nonumber \\%%%%%
 &\Rightarrow 
 \left[
 \Phi(0, \sigma_2 + \beta) 
- 
U^{-1}_{{g}} 
U^{\ }_{{P}} 
U^{\ }_{{g}} \Phi(0, \sigma_2) 
\right]
\mathscr{G}| {C}_{{P}}\rangle 
=0. 
\end{align}
Thus we deduce the following relation:  
\begin{align}
\mathscr{G} |{C}_{{P}}\rangle  
=
|{C}_{{g}\cdot {P} \cdot {g}^{-1}} \rangle.
\end{align}

If the cross-cap condition is invariant under 
$g \in {G}$, 
$U_{P} =U^{-1}_{{g}}U^{\ }_{{P}}U^{\ }_{{g}}$, 
then we may expect that so is the cross-cap state, 
$|{C}_{{P}} \rangle = \mathscr{G} |{C}_{{P}}\rangle$,
classically. 
However this expected invariance may be broken down 
quantum mechanically.
This then signals that the theory is anomalous and should describe 
the edge of a SPT phase defined in one higher dimension.

\section{Bosonic SPT phases}
\label{Bosonic SPT phases}

In this section, 
we apply our strategy above to edge theories of bosonic SPT phases 
consisting of a single component non-chiral boson
(i.e., a two-component chiral boson with $2\times 2$ K-matrix). 
We consider situations where 
parity ($P$) or a combination of parity and charge conjugation ($CP$) is a
part of the full symmetry group, which can potentially protect the edge theory
from gap-opening.  

In Ref.\ \onlinecite{Hsieh2014}, 
SPT phases protected by $P$ or $CP$ symmetry together with a continuous $U(1)$ symmetry    
were studied by using a generalization of Laughlin's argument.
It was found that non-trivial $\mathbb{Z}_2$ bosonic SPT phases can exist
in the presence of the following combinations of
symmetries:
\begin{itemize}
\item
$P \rtimes U(1)_A$ (``spin $U(1)$''), 
\item
$CP \rtimes U(1)_V$ (``charge $U(1)$''),  
\end{itemize}
where the subscript $A/V$ represents the ``axial/vectorial'' nature of the $U(1)$ symmetry. 

In Ref. \onlinecite{Hsieh_CPT}, 
following the spirit of Refs.\ \onlinecite{Lu2012a, Lu2013}, 
a microscopic stability analysis for the bosonic edge theory 
is carried out by enumerating possible gapping potentials 
in the presence of various discrete symmetries. 
It was found that 
the bosonic edge theory can be ingappable (protected)
in the presence of the following symmetries:
\begin{itemize}
\item
$P\times C, P\times T$

\item
$P \times TC$, 
$CP \times T$, 
$T \times C$. 

\end{itemize}
SPTs protected by these symmetries are all classified by $\mathbb{Z}_2$.
In addition, it was also found that there are SPT phases protected by 
\begin{itemize}
\item $T \times C \times P$  
\end{itemize}
The classification of these SPT phases is ${\mathbb Z}^{4}_2$.

In the following,
we will study these SPT phases by using cross-cap states and by identifying quantum anomalies.
As we will show, this analysis reproduces exactly the same classification as 
in Ref.\ \onlinecite{Hsieh_CPT},
and hence it gives us a perspective complimentary to the generalized Laughlin's argument on 
the Klein bottle in the ``loop channel'' picture, and to microscopic stability analysis.  

\subsection{Free compactified boson}
We start from the free boson theory on a spatial ring of circumference $\ell$
defined by the partition function $Z=\int\mathcal{D}[\phi] \exp (iS)$ with the action
\begin{align}
S=\frac{1}{4\pi\alpha'}\int dt \int_{0}^{\ell} dx
\left[
\frac{1}{v}( \partial_{t}\phi )^{2}
-v( \partial_{x}\phi )^{2}
\right],
\end{align}
where 
the spacetime coordinate of the edge theory is denoted by $(t,x)$,
$v$ is the velocity, 
$\alpha'$ is the coupling constant,
and  the $\phi$-field is compactified as
\begin{align}
\phi \sim \phi + 2\pi R,
\end{align}
with the compactification radius $R$. 
The canonical commutation relation is
\begin{align}
\left[
 \phi(x,t), \partial_t {\phi}(x^{\prime},t)
\right]
=
i
2\pi \alpha' v
\sum_{n\in \mathbb{Z}}\delta(x-x^{\prime}-n\ell).
\end{align}
We use the chiral decomposition of the boson field, and introduce the dual field $\theta$ as 
\begin{align}
\phi=\varphi_L+\varphi_R,
\quad
\theta =\varphi_L-\varphi_R. 
\end{align}
The mode expansion of the chiral boson fields is given by ($x^{\pm}=vt\pm x$) 
\begin{align}
\varphi_{L}(x^{+})
&=
x_L
+
\pi \alpha' p_L
\frac{ x^{+}}{\ell}
+
i 
\sqrt{ \frac{\alpha'}{2} }
\sum^{n\neq 0}_{n\in \mathbb{Z}}
 \frac{\alpha_{n}}{{n}}e^{-\frac{2\pi i nx^{+}}{\ell}}
, 
\nonumber \\
\varphi_{R}(x^{-})
&=
x_R
+
\pi \alpha'  p_R
\frac{x^{-}}{\ell}
 +
i \sqrt{ \frac{\alpha'}{2} }
\sum^{n\neq 0}_{n\in \mathbb{Z}}
 \frac{\tilde{\alpha}_{n}}{{n}}e^{-\frac{2\pi i nx^{-}}{\ell}} , 
\end{align}
where  
$ 
\left[\alpha_m, \alpha_{-n}\right] =  
\left[\tilde{\alpha}_m, \tilde{\alpha}_{-n}\right] =  
m\delta_{mn}$ and $ [x_L, p_L]= [x_R, p_R]= {i}$. 
The compactification condition on the boson fields 
implies the allowed momentum eiganvalues are given by  
\begin{align}
&p=\frac{1}{2}\left(p_L + p_R\right) = \frac{k}{R},
\quad
\tilde{p} = \frac{1}{2} \left( p_L - p_R \right) = \frac{R}{\alpha'}w, 
\nonumber \\ 
&p_L = \frac{k}{R}+\frac{R}{\alpha'}w,  
\quad
p_R = \frac{k}{R}-\frac{R}{\alpha'}w,
\end{align}
where $k$ and $w$ are an integer. 
In terms of these momentum eigenvalues, 
the compactification conditions on 
the boson fields are 
\begin{align}
\varphi_L(x+\ell) - \varphi_L(x)&=+\pi \alpha' p_L,
\nonumber\\
\varphi_R(x+\ell) - \varphi_R(x)&= -\pi \alpha' p_R,
\nonumber\\
\phi (x+\ell) - \phi(x)& = \pi \alpha'(p_L-p_R)=2\pi R w,
\nonumber \\
\theta(x+\ell) - \theta(x) &= \pi \alpha'(p_L+p_R) = 2\pi \frac{\alpha'}{R}k. 
\end{align}

The Hilbert space is constructed as a tensor product of 
the bosonic oscillator Fock spaces,  
each of which generated by 
pairs of creation and annihilation operators 
$\{ \alpha_{m}, \alpha_{-m}\}_{m>0}$
and 
$\{\tilde{\alpha}_{m}, \tilde{\alpha}_{-m}\}_{m>0}$, 
and 
the zero mode sector associated to ${x}_{L,R}$ and ${p}_{L,R}$.
We will denote states in the zero mode sector by specifying 
their momentum eigenvalues as
\begin{align}
 |p, \tilde{p}\rangle = | k/R, R w/\alpha'\rangle,
 \quad
 k, w\in \mathbb{Z},  
\end{align}
or more simply as $|k,w\rangle$. 
Alternatively, 
the Fourier transformation of the momentum eigenkets 
defines the ``position'' eigenkets, which we denote by
\begin{align}
 |\phi_0, \theta_0\rangle 
 \quad
 0 <\phi_0 \le 2\pi R,
 \quad
 0< \theta_0 \le 2 \pi \alpha'/R. 
\end{align}
The two basis are related by
\begin{align}
 | p, \tilde{p}\rangle
 =
 \int^{2\pi R}_0 d\phi_0
 \int^{2\pi \alpha'/R}_0 d\theta_0
e^{ -i p \phi_0 -i \tilde{p}\theta_0}
|\phi_0,\theta_0\rangle. 
\end{align}

\subsection{Symmetries}

Various symmetries of the single-component compactified boson theory 
are listed below.

\paragraph{$U(1)\times U(1)$ symmetry}

In the free boson theory, when there is no perturbation,   
there are two conserved $U(1)$ charges, one for each left- and right-moving sector, 
defined by 
\begin{align}
 N_{L,R}&= \int^{\ell}_0  dx\, \partial_x \varphi_{L,R} =
\alpha' \pi p_{L,R},
\end{align}
They satisfy 
\begin{align}
\left[
\varphi_L, N_L
\right]
=
\left[
\varphi_R, N_R
\right]
=
\alpha' \pi {i}.  
\end{align}
Correspondingly to these conserved quantities, 
the free boson theory is invariant under the following 
$U(1)\times U(1)$ symmetry 
\begin{align}
\mathscr{U}_{\delta\phi, \delta \theta}&:
 \phi \to \phi + \delta \phi,
 \quad
\theta \to 
\theta +
\delta \theta, 
\nonumber \\%%%%%
&:\varphi_L 
\to
\varphi_L 
+
\delta \varphi_L, 
\quad 
\varphi_R 
\to
\varphi_R 
+
\delta \varphi_R. 
\end{align}
In terms of the conserved charges, 
the generators of the $U(1)\times U(1)$ transformations are given by 
\begin{align}
 \mathscr{U}^L_{\delta \varphi_L}
& =
e^{i \delta \varphi_L  
 N_{L}/(\alpha' \pi)
}
=
e^{ 
 i \delta \varphi_L 
 p_{L}
}, 
\nonumber \\%%%%%
\mathscr{U}^R _{\delta \varphi_R}
& =
e^{ 
 i \delta \varphi_R
 N_{R}/(\alpha' \pi)
}
=
e^{ 
 i\delta \varphi_R p_{R}
},
\nonumber \\%%%%%
\mathscr{U}_{\delta\phi,\delta\theta} &=
\mathscr{U}^L_{\delta \varphi_L}
\mathscr{U}^R_{\delta \varphi_R}
=
e^{
 i (\delta \phi p + \delta \theta \tilde{p})}. 
\end{align}
Note also that  
$\mathscr{U}_{\delta \phi, \delta \theta}$
acts on the momentum eigenkets as  
\begin{align}
\mathscr{U}_{\delta \phi, \delta \theta}  |p,\tilde{p}\rangle = 
e^{i (p\delta \phi + \tilde{p} \delta\theta)} |p, {\tilde p} \rangle. 
\end{align}

\paragraph{$C$-symmetry}

Particle-hole symmetry or charge conjugation ($C$-symmetry) is unitary and acts on 
the bosonic fields as  
\begin{align}
\mathscr{C}&:\phi \rightarrow - \phi + n_{c}\pi R,
\quad 
\theta \rightarrow -\theta + \frac{m_{c}\pi \alpha'}{R} \nonumber\\ 
&: (x_1, x_2) \rightarrow (x_1, x_2),
\label{csymmetry:phase}
\end{align}
where $(n_{c}, m_{c}) \in \{0,1\}$. 
From these transformation laws of the boson fields, 
we read off the action of $C$-symmetry on the position basis as
\begin{align}
 \mathscr{C}| \phi_0,\theta_0\rangle 
=
e^{i\delta}
 \left|-\phi_{0} + n_{c}\pi R, -\theta_{0} + m_{c}\pi \alpha'/R 
 \right\rangle,  
 \label{phase ambiguity, C-symmetry}
\end{align}
where $e^{i \delta}$ is an unknown phase factor. 
In order to have the relation 
$\mathscr{C}|p, \tilde{p}\rangle \propto |-p, -\tilde{p}\rangle$, 
expected from the commutation relation between $\mathscr{C}$ and $p,\tilde{p}$, 
the phase $\delta$ has to be a constant (independent of $\phi_{0}$ and $\theta_0$). 
The action of $C$-symmetry on the momentum eigenstates is given by 
\begin{align}
\mathscr{C} 
|p, {\tilde p} \rangle 
&= 
e^{i \delta}
e^{-ipn_{c}\pi R - i{\tilde p}\frac{m_{c}\pi \alpha'}{R}} 
|-p, -{\tilde p} \rangle 
\nonumber \\
&= 
e^{i \delta}
e^{-i\pi  kn_{c}  - i \pi w m_{c}} 
|-p, -{\tilde p} \rangle,  
\label{csymmetry:momentum}
\end{align}
where $p=k/R$ and $\tilde{p}=w R/\alpha'$. 
Since $\delta$ is constant, the phase ambiguity is fixed once we specify the action of 
$\mathscr{C}$ on a {\it reference state}, e.g., $|p,\tilde{p}\rangle = |0,0\rangle$.
In our analysis presented below, the reference state and its charge conjugation parity $e^{i\delta}$
plays an important role. 

\paragraph{$T$-symmetry}

Antiunitary time-reversal operator $\mathscr{T}$ acts on the boson fields as 
\begin{align}
\mathscr{T}&: \phi \rightarrow  \phi + n_{T}\pi R,
\quad 
\theta \rightarrow -\theta + \frac{m_{T} \pi \alpha'}{R} \nonumber\\ 
&: (x_1,x_2) \rightarrow (x_1, -x_2),
\end{align}
where $n_{T}, m_{T} \in \{0 ,1 \}$. 

\paragraph{$P$-symmetry}

Parity $\mathscr{P}$ is defined by 
\begin{align}
\mathscr{P}
&: \phi \rightarrow \phi + n_{p}\pi R,
\quad
\theta \rightarrow -\theta + \frac{m_{p}\pi \alpha'}{R}\nonumber\\
&: (x_{1},x_{2}) \rightarrow (\ell -x_{1},x_{2}),
\label{psymmetry}
\end{align}
where $n_{p}, m_{p} \in \{0 ,1 \}$. 

\paragraph{$CP$-symmetry}
The above symmetries can be combined.
For example, $CP$-symmetry is a non-local unitary symmetry,
and defined by
\begin{align}
\mathscr{CP}
&: \phi \rightarrow 
-\phi + n_{cp}\pi R,
\quad
\theta \rightarrow \theta + \frac{m_{cp}\pi \alpha'}{R}\nonumber\\
&: (x_{1},x_{2}) \rightarrow (\ell -x_{1},x_{2}),
\label{cpsymmetry}
\end{align}
where $n_{cp}, m_{cp} \in \{0 ,1 \}$.

\subsection{Crosscap States}

We now move on to the tree channel picture and construct cross-cap states 
by twisting $P$- and $CP$- symmetries.
After rearranging spacetime and exchanging the role of space and time, 
the spacetime is,  
$\mbox{space}\times \mbox{time} = 2\beta \times \ell/2$. 
We parameterize this Euclidean spacetime by $(\sigma_1,\sigma_2)$. 
The original spacetime of $[0, \ell] \times [0, \beta]$ is parameterized by $(x_1, x_2)$ 
(for the mapping between the two spacetime see Fig.\ \ref{fig:xCap}).

\subsubsection{$P$-symmetry}

We first consider the cross-cap state obtained by twisting parity symmetry,
defined in Eq.\ (\ref{psymmetry}).
The corresponding cross-cap states are
defined by
\begin{align}
&\bigg[ \phi(\sigma_{2}) - \phi(\sigma_{2} + \beta) - n_{p}\pi R \bigg]|C_{p} (n_{p}, m_{p}) \rangle =0,
\nonumber\\
&\bigg[\theta(\sigma_{2}) +\theta (\sigma_{2} + \beta) - \frac{m_{p}\pi \alpha'}{R}\bigg]|C_{p} (n_{p}, m_{p}) \rangle =0.
\label{xcapcondition:p}
\end{align}
%We mode-expand the field variables only upto the zero modes which turns out to be enough to diagnose the SPT phases. 
By mode-expansion,  
the cross-cap condition (\ref{xcapcondition:p}) translates into the corresponding condition for each mode. 
To find an explicit form of the cross-cap states, we first focus on the zero mode sector of the boson fields:  
\begin{align}
\phi(\sigma_{2}) &= x_{L} + x_{R} + \frac{\pi \alpha' {\tilde p} \sigma_{2}}{\beta}+\cdots,
\nonumber\\
\theta(\sigma_{2}) &= x_{L} - x_{R} + \frac{\pi \alpha' p \sigma_{2}}{\beta} +\cdots. 
\end{align}
Within the zero mode sector, we solve the cross-cap conditions \eqref{xcapcondition:p}. 
The first condition \eqref{xcapcondition:p} can be reduced to
\begin{align}
%&\phi(\sigma_{2}) - \phi(\sigma_{2} + \beta) - n_{p}\pi R = 0 \text{ mod } 2\pi R \nonumber\\
&\quad \pi \alpha' {\tilde p} -n_{p}\pi R = 0 \text{ mod } 2\pi R \nonumber\\ 
&\Rightarrow {\tilde p} = \frac{R}{\alpha'} (2N + n_{p}),\quad N \in {\mathbb Z}. 
\label{xcapsolution1:p}
\end{align}
Similarly, the second condition can be solved as: 
\begin{align}
%&\theta(\sigma_{2}) +\theta (\sigma_{2} + \beta) - \frac{m_{p}\pi \alpha'}{R} =0 \text{ mod } \frac{2\pi \alpha'}{R} \nonumber\\ 
&2(x_{L}- x_{R}) + \frac{2\pi \alpha' p \sigma_2}{\beta} + \pi \alpha' p = \frac{m_{p}\pi \alpha'}{R} \text{ mod } \frac{2\pi \alpha'}{R} \nonumber\\ 
& \Rightarrow p=0,
\quad  (x_{L} - x_{R}) = (2{\tilde N} + m_{p}) \frac{\pi \alpha'}{2R}, ~ {\tilde N} \in {\mathbb Z}.  
\label{xcapsolution2:p}
\end{align}
Solving the conditions \eqref{xcapsolution1:p} and \eqref{xcapsolution2:p},
we find the cross-cap state 
$|C_{p} (n_{p}, m_{p}) \rangle$
in terms of the momentum eigenket $\{ |p, {\tilde p} \rangle \}$ as
\begin{equation}
|C_{p} (n_{p}, m_{p}) \rangle = \sum_{N \in {\mathbb Z}} (-1)^{m_{p}N} |0, 2N+n_{p} \rangle. 
\label{xcapstate:p}
\end{equation}
The full cross-cap state is obtained by including the parts related to the oscillatory modes:
\begin{align}
\sqrt{R\sqrt{2} }  
\exp \left[
-\sum_{n=1}^{\infty} \frac{(-1)^n}{n} \alpha_{-n}\tilde{\alpha}_{-n} 
\right]
|C_p (n_p, m_p)\rangle. 
\end{align}
For our purpose of diagnosing SPT phases, however, it turns out that it is enough
to focus on the zero-mode sector.

\subsubsection{$CP$-symmetry}

Next we consider $CP$-symmetry.
The corresponding cross-cap conditions are given by 
\begin{align}
&\bigg[ \phi(\sigma_{2}) + \phi(\sigma_{2} + \beta) - n_{cp}\pi R \bigg]|C_{cp} (n_{cp}, m_{cp}) \rangle =0,
\nonumber\\
&\bigg[\theta(\sigma_{2}) -\theta (\sigma_{2} + \beta) - \frac{m_{cp}\pi \alpha'}{R}\bigg]|C_{cp} (n_{cp}, m_{cp}) \rangle =0.
\nonumber\\
\label{xcapcondition:cp}
\end{align}
By solving these conditions within the zero mode sector,
we obtain, as the cross-cap state,
\begin{equation}
|C_{cp} (n_{cp}, m_{cp}) \rangle = \sum_{N \in {\mathbb Z}} (-1)^{n_{cp}N} |2N+m_{cp},0 \rangle. 
\label{xcapstate:cp}
\end{equation}

\subsection{Diagnosis of SPT phases}
\label{diagnosis}

\subsubsection{$P \rtimes U(1)_A$ and $CP \rtimes U(1)_V$ symmetries}

We start with the case of the symmetry group $P\rtimes U(1)_A$,  
which has been studied in Ref.\ \onlinecite{Hsieh2014}
by using a generalization of Laughln's gauge argument. 
In the generalized Laughlin's argument,
the edge theory is put on an unoriented  spacetime, such as the Klein bottle,  
and then the invariance under flux threading 
(a large gauge transformation of $U(1)_A$ symmetry) 
of the partition function of the edge theory is investigated. 
More specifically, 
the analysis in Ref.\ \onlinecite{Hsieh2014} is performed  
in the loop-channel channel picture. 
In the following, we will reproduce this result in terms of the tree-channel calculations,
i.e., by using the cross-cap state.

The relevant cross-cap state is $|C_p (n_p,0)\rangle$ presented in Eq.\ (\ref{xcapstate:p}).
We set $m_p =0$ since $n_{p} \in \{0,1\}, m_p =0$ is enough to classify the $P\rtimes U(1)_A$-symmetric SPT phases
according to Ref.\ \onlinecite{Hsieh_CPT}. 
However, it is straightforward to generalize the analysis below to more general sets of $\{n_p, m_p \}$. The $U(1)_{A}$ symmetry is generated by 
$\mathscr{U}^A_{\delta \theta} = \exp(i{\tilde p} \delta \theta)$: 
\begin{align}
{\mathscr U}^{A}_{\delta \theta}&: 
\theta \rightarrow \theta+\delta \theta,
\quad 
\phi \rightarrow \phi.  
\end{align}
Let us act with $\mathscr{U}^A_{\delta\theta}$ on the cross-cap condition (\ref{xcapcondition:p}).
The cross-cap condition for $\phi$ is trivially invariant under $\mathscr{U}^A_{\delta\theta}$. 
For the condition written in terms of the dual field $\theta$ (with $m_{p}=0$), 
\begin{align}
&
\mathscr{U}_{\delta\theta}^{A}
\left\{
\theta(\sigma_{2}) +\theta (\sigma_{2} + \beta) 
%- \frac{m_{p}\pi \alpha'}{R}
\right\}
(\mathscr{U}_{\delta\theta}^{A})^{-1}
\mathscr{U}_{\delta\theta}^{A}
|C_{p} (n_{p}, m_{p}) \rangle =0
\nonumber \\
\Rightarrow 
&
\left\{
\theta(\sigma_{2}) +\theta (\sigma_{2} + \beta) 
%- \frac{m_{p}\pi \alpha'}{R}
+2\delta \theta 
\right\}
\mathscr{U}^{A}_{\delta\theta}
|C_{p} (n_{p}, m_{p}) \rangle =0.
\end{align}
Thus, the cross-cap condition is transformed into 
\begin{align}
&
\quad 
\theta(\sigma_{2}) +\theta (\sigma_{2} + \beta) 
%- \frac{m_{p}\pi \alpha'}{R}
=0
\nonumber \\
& 
\Rightarrow 
\theta(\sigma_{2}) +\theta (\sigma_{2} + \beta) 
%- \frac{m_{p}\pi \alpha'}{R}
+2\delta \theta 
=0,
\end{align}
which is invariant when
\begin{align}
2 \delta \theta & = 0 \mod \frac{2\pi \alpha'}{R}. 
\end{align}
Thus, at least classically, we expect the theory to be invariant when $\delta \theta  =\pi \alpha'/R$.  
%from $0$ to $\pi \alpha'/R$. 
On the other hand, this invariance may not be maintained at the quantum level. 
The cross-cap state may not be invariant under $\mathscr{U}^A_{\delta \theta = \pi \alpha'/R}$ 
and can pick up an anomalous phase;
one can easily check  
\begin{align}
{\mathscr U}^A_{\pi \alpha'/R} |C_p (n_{p}, 0) \rangle = (-1)^{n_{p}}| C_p (n_{p},0) \rangle.
\end{align}
Thus the edge theory, when enforced parity symmetry with $n_{p}=1$,
is anomalous for $U(1)_A$ symmetry. 
This invariance/non-invariance is equivalent to the 
invariance/non-invariance of the Klein bottle partition function under 
large gauge transformations discussed in Ref.\ \onlinecite{Hsieh2014}.
Observe also that the anomalous phase here is ${\mathbb Z}_{2}$-valued (i.e., a sign), 
and this implies that the classification is ${\mathbb Z}_{2}$ since 
the two copies of the theory is trivial. 
This agrees with the loop channel calculation.\cite{Hsieh2014, Hsieh_CPT}

SPT phases protected by $CP \rtimes U(1)_V$
can be discussed in the same manner as  $P \rtimes U(1)_A$. 
The cross-cap state obtained by twisting $CP$ is given by
$|C_{cp} (0, m_{cp})\rangle$ in Eq.\ (\ref{xcapstate:cp}).  
%\begin{equation}
%|C_{cp}, (0,m_{cp}) \rangle = \sum_{\omega \in {\mathbb Z}} |2\omega + m_{cp},0\rangle, 
%\quad m_{cp}= 0, 1.
%\end{equation}
The $U(1)_{V}$ symmetry is generated by 
${\mathscr U}^{V}_{\delta \phi} = \exp(ip \delta \phi)$ as
\begin{align}
{\mathscr U}^{V}_{\delta \phi}&: \theta \rightarrow \theta,
\quad 
\phi \rightarrow \phi + \delta \phi. 
\end{align}
Under this symmetry action, the cross-cap condition \eqref{xcapcondition:cp} with $n_{cp}=0$ 
is invariant when  
\begin{align}
2\delta \phi & = 0 \mod 2\pi R. 
\end{align}
On the other hand, we find 
\begin{align}
{\mathscr U}^{V}_{\delta \phi = \pi R} |C_{cp} (0, m_{cp}) \rangle = (-1)^{m_{cp}}| C_{cp} (0,m_{cp}) \rangle.
\end{align}
Thus the edge theory with parity symmetry with $m_{cp}=1$ and $U(1)_V$ is anomalous.
\cite{Hsieh2014, Hsieh_CPT} 
Here the anomalous phase is the ${\mathbb Z}_{2}$ sign, 
and thus classification is ${\mathbb Z}_{2}$ as in the previous case.

The above analysis in terms of the cross-cap states is a reformulation of Ref. \onlinecite{Hsieh2014};
while in Ref. \onlinecite{Hsieh2014} the Klein bottle partition functions are computed in the loop channel picture,
here we have considered and studied the cross-cap states in the tree channel. 
A few technical remarks are in order. 
(i) 
We have considered adiabatic transformations of the cross-cap states
by acting with 
$\mathscr{U}^A_{\delta\theta}$ or $\mathscr{U}^V_{\delta\phi}$
and by continuously changing   
$\delta \theta$ or $\delta \phi$, respectively.
In terms of the original, loop channel picture, 
these twisting parameters should appear as 
a twisting angle in twisting boundary conditions in the spatial direction. 
This can be seen explicitly as follows. 
By using the formula
$
\mathscr{G}|C_P\rangle = |C_{ g P g^{-1}}\rangle, 
$
the partition function in the tree channel can be written as,
\begin{align}
Z^K &= 
\langle C_{P'} | 
e^{ -\frac{\ell}{2} H_{\mathrm{tree}} } 
\mathscr{U}^A_{\delta \theta} 
|C_P\rangle  
\nonumber \\
  &= \langle C_{P'} | 
  e^{-\frac{\ell}{2} H_{\mathrm{tree} } } 
  | C_{ U^A_{\delta \theta} \cdot P \cdot U^{A}_{-\delta \theta}} \rangle.  
\end{align}
This can be written in the loop channel as 
\begin{align}
 Z^K &= 
  \mathrm{Tr}^{\ }_{ P' U^A_{\delta\theta} P^{-1} U^{A}_{-\delta \theta} }\, e^{-\beta H_{\mathrm{loop}} }  
\end{align}
By noting 
$U^A_{\delta \theta} P^{-1} U^A_{-\delta\theta}= P^{-1} U^A_{-2\delta \theta}$,
and taking $P'=P$,  
\begin{align}
 Z^K = \mathrm{Tr}_{U^{A}_{-2\delta\theta}}
  e^{-\beta H_{\mathrm{loop} }},   
\end{align}
i.e., $2\delta \theta$ (not $\delta \theta$) 
appears, in the loop channel, as a twisting angle for a spatial boundary condition. 
When $2\delta \theta$ is an integer multiple of $2\pi$, 
the twisted system is large gauge equivalent to the system without twisting.

(ii) 
The current analysis in the tree-channel picture has a few advantages over the loop-channel calculations.
First of all, 
in Ref.\ \onlinecite{Hsieh2014},
we rely heavily on the existence of a continuous $U(1)$ symmetry
(either $U(1)_A$ or $U(1)_V$) as in Laughlin's thought experiment in the quantum Hall effect. 
Therefore, it is not entirely obvious how the methodology in Ref.\ \onlinecite{Hsieh2014}
can be generalized to SPT phases which lack continuous ($U(1)$) symmetries.
The reformulation in the tree channel and in term of cross-cap states, however, 
indicates a natural way to deal with SPTs without $U(1)$ symmetry.
In fact, the procedure described above can be generalized to cases without 
$U(1)$ symmetry. 
(See the following sections.)
Second, 
in the current reformulation in the tree channel,
there is no need to compute the full partition function,
although it is possible to compute the partition function by using cross-cap states.

\subsubsection{$P \times C$ symmetry}

To demonstrate that our methodology works in the absence of a continuous $U(1)$ symmetry, 
we now consider SPT phases protected by $P \times C$. 
Here the relevant cross-cap state is $|C_p (n_{p},m_{p}) \rangle$ in Eq.\ \eqref{xcapstate:p}.  
%\begin{equation}
%|C_p, (n_{p},m_{p}) \rangle = \sum_{\omega \in {\mathbb Z}} (-1)^{\omega m_{p}}|0, 2\omega + n_{p}\rangle 
%\end{equation}
%in which $n_{p}, m_{p} \in \{ 0, 1\}$. 
We consider $C$-symmetry \eqref{csymmetry:phase} with $n_{c}=m_{c}=0$. 
One can check easily that under the action of $C$-symmetry, the cross-cap condition \eqref{xcapcondition:p} is invariant. 
On the other hand, the cross-cap state may not, as one can check explicitly as 
\begin{align}
\mathscr{C} |C_p (n_{p}, m_{p}) \rangle = e^{i\delta} (-1)^{n_{p}m_{p}}| C_p (n_{p},m_{p}) \rangle,
\label{1}
\end{align}
by using Eq.\ \eqref{csymmetry:momentum}. 
To judge if the cross-cap state is anomalous or not,
we need to know the overall phase $e^{i \delta}$, which originates from 
our ignorance on the charge conjugation parity of the zero mode sector.
Depending on our choice of $\delta$, 
either one of phases with 
$(n_p,m_p)=(0,0),(0,1), (1,0)$ or $(n_p,m_p)=(1,1)$ is anomalous.
A natural choice would be $e^{i\delta}=1$,
as this means we assign the charge conjugation parity 
$+1$ to the reference state $|p,\tilde{p}\rangle = |0,0\rangle$. 
With this choice, the edge theory with the symmetry group $P\times C$ 
is anomalous when $(n_{p},m_{p})=(1,1)$:
this result is consistent with the microscopic analysis given in Ref. \onlinecite{Hsieh_CPT}.
The phase acquired by the cross-cap state under the action of $C$-symmetry is  ${\mathbb Z}_{2}$ (i.e., sign), 
and hence the classification is ${\mathbb Z}_{2}$. 
Furthermore, one may twist $CP$ symmetry, 
which can be generated by the multiplication of $C$ and $P$, i.e., $CP = C \times P$, 
to form the cross-cap state $|C_{cp}, (n_p, m_p) \rangle$ and then study the action of $C$ on the cross-cap:
\begin{align}
\mathscr{C} |C_{cp} (n_{p}, m_{p}) \rangle = e^{i\delta'} (-1)^{n_{p}m_{p}}| C_{cp} (n_{p},m_{p}) \rangle, 
\end{align}
where $e^{i\delta'}$ is an overall phase factor, 
which, as in Eq.\ \eqref{1},  cannot be determined. 
This again generates the same classification as Eq.\ \eqref{1}.

This situation should be contrasted with the cases with
a continuous $U(1)$ symmetry; in the latter case, 
one can compare the phase 
acquired by the cross-cap state when acting with 
$\mathscr{U}^V_{\delta\theta}$
and
$\mathscr{U}^V_{\delta\theta+\pi R}$; 
One can follow the evolution of $\mathscr{U}^V_{\delta\theta}|C_{cp} (n_{cp}, m_{cp})\rangle$
adiabatically.  
In the present case, because of the discrete nature of $C$-transformation,
such comparison is not possible.

\subsubsection{$P\times T$ symmetry}

We now consider the cases where we have both $P/CP$ and $T/CT$ symmetries.
As in the case of $P\times C$, we can first twist $P/CP$ to obtain the corresponding cross-cap state.
The action of the remaining symmetry, $T/CT$, on the cross-cap state can be studied. 
However, unlike $C$-symmetry, which is a non-spatial unitary symmetry, 
how $T/CT$ symmetry acts in the tree-channel is non-trivial. 
We illustrate this point for the case of $P\times T$ symmetry. 

Let us consider the bosonic edge theory 
in the presence of $P$-symmetry with $m_{p}=0$ 
and $T$-symmetry with $n_T=0$.
We twist $P$-symmetry and consider the corresponding cross-cap state 
$|C_p (n_p,0)\rangle$ in Eq.\ \eqref{xcapstate:p}. 
One can check the cross-cap condition is left invariant under $T$-symmetry:
the fields obey, in the loop channel, the following boundary conditions:
\begin{align}
 \phi(x_1, x_2+\beta) &\equiv \phi(\ell-x_1, x_2) + n_p \pi R, 
\nonumber \\
 \phi(x_1+\ell, x_2) &\equiv \phi(x_1, x_2),  
 \nonumber \\%%%%%
 \theta(x_1, x_2+\beta) &\equiv -\theta(\ell-x_1, x_2), 
 \nonumber \\
 \theta(x_1+\ell, x_2) &\equiv \theta(x_1, x_2).  
\label{62}
\end{align}
Under time-reversal, these conditions are transformed into: 
\begin{align}
& \phi(x_1, -x_2) \equiv \phi(\ell-x_1, \beta -x_2) + n_p \pi R, 
 \nonumber \\%%%%%
& \phi(x_1+\ell, \beta-x_2) \equiv \phi(x_1, \beta-x_2),  
\nonumber \\ 
& -\theta(x_1, -x_2)+m_T\frac{\pi \alpha'}{R} \equiv 
 \theta(\ell-x_1, \beta-x_2)-m_T \frac{\pi \alpha'}{R}, 
 \nonumber \\
& -\theta(x_1+\ell, \beta-x_2)+m_T \frac{\pi\alpha'}{R} \equiv 
 -\theta(x_1, \beta-x_2) + m_T \frac{\pi\alpha'}{R}.    
\label{63}
\end{align}
For example, we can derive the first line of Eq.\eqref{62} from the first line of Eq.\eqref{63} by following steps. 
\begin{align}
&\phi(x_1, x_2+\beta) \equiv \phi(\ell-x_1, x_2) + n_p \pi R, \nonumber\\ 
&\leftrightarrow \phi(x_1, y_2) \equiv \phi(\ell-x_1, y_2 - \beta) + n_p \pi R, \nonumber\\ 
&\to \phi(x_1, -y_2) \equiv \phi(\ell-x_1, -y_2 + \beta) + n_p \pi R, \nonumber\\ 
&\leftrightarrow \phi(x_1, -x_2) \equiv \phi(\ell-x_1, \beta - x_2) + n_p \pi R,
\end{align}
where we renamed the variable $y_2 = x_2 + \beta$ inbetween the first and the second lines, and we acted $T$-symmetry inbetween the second and the third lines. With the compactification conditions, these are equivalent to 
\begin{align}
 \phi(x_1,  -x_2) &\equiv \phi(\ell-x_1, \beta-x_2) + n_p \pi R, 
 \nonumber \\%%%%%
 \phi(x_1+\ell, \beta-x_2) &\equiv \phi(x_1, \beta-x_2),  
\nonumber \\ 
 \theta(x_1, -x_2) &\equiv -\theta(\ell-x_1, \beta-x_2), 
 \nonumber \\
 \theta(x_1+\ell, \beta-x_2) &\equiv 
 \theta(x_1, \beta-x_2),    
\end{align}
which, with mere relabeling $\beta-x_2\to x_2$, 
are equivalent to the original boundary conditions. 

%We now analyze the action of $\mathscr{T}$ on the cross-cap state. 
Since $T$-transformation leaves the $P$-twisted boundary condition invariant, 
we expect the cross-cap state is also invariant under $T$-transformation,
at least up to a phase factor. 
(Recall that we are after this possible anomalous phase of the cross-cap state.) 
To compute this phase, we need to know how $T$-transformation looks like 
in the tree channel. 
In the tree channel, $T$-transformation should look like a parity transformation 
(in fact, $CP$-transformation, since $T$ flips the sign of $\phi$):  
\begin{align}
 \phi(\sigma_1,\sigma_2) &\to -\phi(\sigma_1, 2\beta-\sigma_2),
 \nonumber \\%%%%%
 \theta(\sigma_1,\sigma_2)&\to \theta(\sigma_1, 2\beta-\sigma_2) 
 +m_T \frac{\pi \alpha'}{R}. 
\end{align}
This $CP$-transformation in the tree-channel picture, obtained from $T$-transformation in the loop-channel picture,
is denoted by $\widetilde{CP}$ in the following. 
If the phase field $\phi$ and its dual $\theta$ obey the standard commutation 
relation, this $\widetilde{CP}$-transformation must be unitary, 
i.e., it must preserve, in particular, the Heisenberg algebra obeyed by 
the zero modes, $[x_L, p_L]=[x_R,p_R]=i$.
If it were defined in the original coordinates $(x_1,x_2)$, 
this unitary $\widetilde{CP}$-transformation is CPT-dual of $T$-symmetry.  

Since $T$-transformation in the loop channel 
preserves the boundary condition, this should be so in tree channel as well. 
The cross-cap conditions
\begin{align}
 &
\left[
 \phi(\sigma_2) - \phi(\sigma_2 + \beta) - n_p \pi R
\right]
| C_p (n_p,0) \rangle
=
0, 
\nonumber \\%%%%%
 &
\left[
 \theta(\sigma_2) + \theta(\sigma_2 + \beta) 
\right]
| C_p (n_p,0) \rangle
=
0, 
\end{align}
are transformed into, by the $\widetilde{CP}$ transformation, 
\begin{align}
 &
\left[
 -\phi(2\beta-\sigma_2) + \phi(\beta-\sigma_2 ) - n_p \pi R
\right]
\widetilde{\mathscr{CP}}| C_p (n_p,0) \rangle
=
0, 
\nonumber \\%%%%%
 &
\Big[
 \theta(2\beta-\sigma_2) 
 + m_T \frac{\pi \alpha'}{R}
 \nonumber \\
&\qquad   
 + \theta(\beta-\sigma_2) 
 + m_T \frac{\pi \alpha'}{R}
\Big]
\widetilde{\mathscr{CP}}| C_p (n_p,0) \rangle
=
0.
\end{align}
Due to the compactification condition,
these cross-cap conditions are equivalent to the original conditions. 
While the crosscap conditions are preserved by $\widetilde{CP}$, 
the cross-cap state $|C_p (n_p ,0) \rangle$ \eqref{xcapstate:p} is not invariant 
when $n_p =m_T = 1$,  
\beq
\widetilde{\mathscr{CP}}|C_p (n_p,0) \rangle = (-1)^{m_T n_p} |C_p (n_p,0) \rangle. 
\eeq
Thus $\widetilde{CP}$ or the original time-reversal symmetry is anomalous. 
This result is consistent with the gapping potential analysis in Ref.\ \onlinecite{Hsieh_CPT}.

In Appendix \ref{Bosonic SPT phases app}, 
we present identification of quantum anomalies using cross-cap states 
for  other bosonic SPT phases studied in  Ref.\ \onlinecite{Hsieh_CPT}.

\section{Real fermionic SPTs}
\label{Real fermionic SPTs}

In this section, 
we discuss edge theories of (2+1)d topological superconductors in the presence of discrete symmetries. 
In particular, we consider topological superconductors with reflection symmetry in 
symmetry classes D+R$_{+}$ and BDI + R$_{++}$, and their CPT partners 
discussed in Refs.\ \onlinecite{Chiu2013, Morimoto2013, Shiozaki2014},
At quadratic level, i.e., in the absence of interactions,   
topological classification for these symmetry classes 
are found to be $\mathbb{Z}_2$ and $\mathbb{Z}$, respectively.  
For the latter, it was found in Ref.\ \onlinecite{yao2013interaction}, the $\mathbb{Z}$ classification collapses into
$\mathbb{Z}_8$ in the presence of interactions. 
Our discussion in this section can trivially be extended to other fermionic SPT phases
supporting complex (Dirac) fermion edge modes.  

\subsection{Majorana fermion edge states}

Consider an edge theory of a (topological) superconductor 
consisting of $N_f$ flavors of non-chiral real (Majorana) fermions
described by the Hamiltonian 
\begin{align}
H
&=
\sum^{N_f}_{a=1}
\int^{\ell}_0dx\, 
\left[
\psi^a_L ( -v i\partial_x) \psi^a_L
+
\psi^a_R ( + v i \partial_x) \psi^a_R
\right]. 
\label{fermionic edge theory}
\end{align}
The fermi velocity $v$ is set to be identical for all fermion flavors 
for simplicity. 
The fermion fields obey the canonical anticommutation relations
\begin{align}
\{ \psi^a_L(x), \psi^b_L(x')\} &= 2\pi \delta^{ab} \sum_{m\in \mathbb{Z}} \delta(x-x'+\ell m),
\nonumber \\
\{ \psi^a_R(x), \psi^b_R(x')\} &= 2\pi \delta^{ab} \sum_{m\in \mathbb{Z}} \delta(x-x'+\ell m).
\end{align}
The (1+1)d non-chiral fermionic edge theory (\ref{fermionic edge theory})
can be realized at the edge of (2+1)d topological superconductors
in symmetry classes DIII, D+R$_{+}$ and BDI + R$_{++}$.  
The fermionic edge theory (\ref{fermionic edge theory})
is invariant under the following three symmetries:

(i)
The fermion number parity conservation,
where the fermion parity operator is given by
\begin{align}
\mathscr{G}_f = 
(-1)^{F},
\quad
F =\sum^{N_f}_{a=1} F_a, 
\end{align}
where $F_a$ is the total fermion number operator for the $a$-th flavor, 
\begin{align}
 F_a = \frac{1}{2\pi} \int^{\ell}_0dx\,  i \psi^a_L \psi^a_R. 
\end{align}
The fermion number parity conservation is the most fundamental symmetry of 
any fermionic system. 
In the following, the fermion number parity conservation 
is assumed to be always unbroken
(at least classically -- it may however be anomalous).

(ii) 
Time-reversal symmetry.
We consider two-kinds of time-reversal:
one which squares to $+1$:
\begin{align}
&
\mathscr{T} \psi_L(t,x) \mathscr{T}^{-1} = \psi_R(-t,x),
\nonumber \\%%%%%
&
\mathscr{T} \psi_R(t,x) \mathscr{T}^{-1} = \psi_L(-t,x), 
\nonumber \\
& \quad \mathscr{T}^2=1,
\quad 
\mathscr{T}i \mathscr{T}^{-1} = -i, 
\end{align}
and  
the other which squares to $-1$:
\begin{align}
&
\mathscr{T} \psi_L(t,x) \mathscr{T}^{-1} = \psi_R(-t,x),
\nonumber \\%%%%%
&
\mathscr{T} \psi_R(t,x) \mathscr{T}^{-1} = -\psi_L(-t,x), 
\nonumber \\
& \quad \mathscr{T}^2=\mathscr{G}_f,
\quad 
\mathscr{T}i \mathscr{T}^{-1} = -i. 
\end{align}
(The flavor index is suppressed.)

(iii) 
Parity symmetries. 
Similarly to time-reversal, 
we consider two-kinds of parities: 
one which squares to $+1$:
\begin{align}
&\mathscr{P} \psi_L(t,x) \mathscr{P}^{-1} = \psi_R(t,\ell-x),
\nonumber \\%%%%%
&\mathscr{P} \psi_R(t,x) \mathscr{P}^{-1} = \psi_L(t,\ell-x), 
\nonumber \\
&\quad 
U_P = \sigma_x, \qquad \mathscr{P}^2=1,   
\end{align}
and 
the other which squares to $-1$:
\begin{align}
&\mathscr{P} \psi_L(t,x) \mathscr{P}^{-1} = \psi_R(t,\ell-x),
\nonumber \\%%%%%
&\mathscr{P} \psi_R(t,x) \mathscr{P}^{-1} = -\psi_L(t,\ell-x).  
\nonumber \\
& \quad 
U_P = i \sigma_y, \qquad \mathscr{P}^2 = -1. 
\end{align}
Note that the phase of parity (and hence parity squared) is arbitrary 
for systems of complex fermions whereas 
there is no such arbitrariness for real fermions. 

Observe that 
$\mathscr{T}^2=-1$ and $\mathscr{P}^2=1$ are CPT conjugate to each other. 
Similarly,  
$\mathscr{T}^2=1$ and $\mathscr{P}^2=-1$ are CPT conjugate to each other. 
(These can easily be inferred from 
the fact that they both prohibit the same kind of masses.)
In the absence of time-reversal and parity symmetries, 
the non-chiral fermion edge state can easily be gapped by adding a mass term.
To realize a non-trivial SPT phase, 
we need to impose $\mathscr{T}$, $\mathscr{P}$, or both $\mathscr{T}$ and $\mathscr{P}$
(see Table \ref{table}).
Imposing $\mathscr{T}^2=-1$ alone leads to 
$\mathbb{Z}_2$ topological classification (class DIII)
while imposing $\mathscr{T}^2=1$ alone does not give rise to a SPT. 
Similarly, imposing $\mathscr{P}^2=1$ alone leads to 
$\mathbb{Z}_2$ topological classification (class D + R$_+$)
while imposing $\mathscr{P}^2=-1$ alone does not give rise to a SPT (class D+R$_-$). 
Imposing $\mathscr{T}^{2}=-1$ together with $\mathscr{P}^2=-1$ (class DIII + R$_{--}$) 
gives rise to $\mathbb{Z}_8$ topological distinction. 
Similarly, its CPT conjugate,
$\mathscr{P}^2=1$ together with $\mathscr{T}^2=+1$ (class BDI + R$_{++}$) 
gives rise to $\mathbb{Z}_8$ topological distinction. 
In the following, our task is to understand
these $\mathbb{Z}_2$ (class D+R$_{+}$) and 
$\mathbb{Z}_8$ (class BDI+R$_{++}$) classifications of SPTs in terms of 
quantum anomalies. 
We will apply our formalism using cross-cap states.

\begin{table}
 \begin{tabular}{c|l|l|c}
 Class & $T$ & $P$ & Classification  
 \\ \hline
DIII & $\mathscr{T}^2=\mathscr{G}_f$ & None & $\mathbb{Z}_2$ \\
BDI & $\mathscr{T}^2=1$ & None & 0 \\
D+R$_{+}$ & None & $\mathscr{P}^2=1$ & $\mathbb{Z}_2$ \\
D+R$_{-}$ & None & $\mathscr{P}^2=-1$ & 0 \\
DIII+R$_{--}$ & $\mathscr{T}^2=\mathscr{G}_f$ & $\mathscr{P}^2=-1$ & $\mathbb{Z}\to \mathbb{Z}_8$ \\
%DIII+R$_{++}$ & $\mathscr{T}^2=\mathscr{G}_f$ & $\mathscr{P}^2=-1$ & $\mathbb{Z}_2$ \\
BDI+R$_{++}$ & $\mathscr{T}^2=1$ & $\mathscr{P}^2=1$ & $\mathbb{Z}\to \mathbb{Z}_8$ \\
\end{tabular}
\caption{
Edge theories of (2+1)d topological superconductors with various time-reversal and parity symmetries
studied in Sec.\ \ref{Real fermionic SPTs}. 
\label{table}
}
\end{table}

\subsection{Cross-cap states}

\paragraph{Cross-cap condition}
In both symmetry classes D+R$_{+}$ and DIII+R$_{--}$,
parity $\mathscr{P}$ and fermion number parity $G_f$ are conserved.   
Hence, theories in these symmetry classes can be twisted by 
parity $\mathscr{P}$, and the combination of fermion number parity and spatial parity $G_f \mathscr{P}$. 
In the fermionic edge theory (\ref{fermionic edge theory}), 
twisting boundary conditions in time direction by $\mathscr{P}$ or $\mathscr{G}_f \mathscr{P}$
leads to the following conditions on the fermion fields: 
\begin{align}
\psi_L(x_1, x_2+\beta) =\eta_1 \psi_R(\ell-x_1, x_2), 
\nonumber \\
\psi_R(x_1, x_2+\beta) = \eta_2 \psi_L(\ell-x_1, x_2). 
\end{align}
Here, $\eta_{1,2}$ are given by 
\begin{align}
 (\eta_1, \eta_2) = 
 \left\{
 \begin{array}{ll}
 (+,+)& \mbox{twisting by $\mathscr{P}$ with $\mathscr{P}^2=1$} 
 \\
 (-,-)& \mbox{twisting by $\mathscr{G}_f \mathscr{P}$ with $\mathscr{P}^2=1$} 
 \\
 (+,-)& \mbox{twisting by $\mathscr{P}$ with $\mathscr{P}^2=-1$} 
 \\
 (-,+)& \mbox{twisting by $\mathscr{G}_f \mathscr{P}$ with $\mathscr{P}^2=-1$} 
 \end{array}
 \right.
\end{align}
In the rearranged geometry (see Fig.\ \ref{fig:xCap}),  
these twisted boundary conditions are given by 
\begin{align}
\psi_L(\sigma_1, \sigma_2) &=\eta_1 \psi_R(\sigma_1, \sigma_2+\beta), 
\nonumber \\
\psi_R(\sigma_1, \sigma_2) &= \eta_2 \psi_L(\sigma_1, \sigma_2+\beta). 
\end{align}
By making a 90 degree rotation in the $(\sigma_1, \sigma_2)$ plane,
$(\sigma_1, \sigma_2) \to (\sigma_1', \sigma_2') = (\sigma_2, -\sigma_1)$, 
we exchange the role of time and space coordinates and regard $\sigma_1$ direction as a fictitious time direction.
The fermion fields $\psi^{\prime}_{L,R}$ with respect to the rotated coordinate system $(\sigma_1', \sigma_2')$
are related to the original fermion fields as 
\begin{align}
 \psi^{\prime}_L = e^{i\pi/4} \psi_L, 
 \quad
 \psi^{\prime}_R = e^{-i\pi/4} \psi_R.  
\end{align}
The cross-cap states are defined by the cross-cap conditions
\begin{align}
  \left[
   \psi_L (\sigma_2)- i \eta_1 \psi_R (\sigma_2+\beta)
  \right]
  |C (\eta_1,\eta_2) \rangle &=
  0, 
\nonumber \\
 \left[
  \psi_R (\sigma_2)+ i \eta_2 \psi_L (\sigma_2+\beta)
 \right]
 |C (\eta_1, \eta_2) \rangle &=0, 
\end{align}
where for notational simplicity we have removed 
${ }^{\prime}$ and simply 
write $\psi^{\prime}_{L,R}\to \psi_{L,R}$
and $\sigma^{\prime}_{1,2}\to \sigma_{1,2}$. 

A crucial observation is that 
for the cross-cap conditions generated by $\mathscr{P}$ with $\mathscr{P}^2=+1$,  
there are Majorana fermion zero modes in the mode expansion of the
fermion fields, 
while there are no Majorana zero modes when $\mathscr{P}^2=-1$. 
This follows from the general formula (\ref{cross cap formula 1}),
which suggests that when $\mathscr{P}^2=1(-1)$ the fermion fields obey periodic (anti-periodic) boundary condition.
(This can also be understood by first assuming 
the existence of zero modes, $\psi_{L0}$ and $\psi_{R0}$ for 
the left- and right-moving fermion fields, respectively.
Then, the cross-cap condition tells us
$\psi_{L0} - i\eta_1 \psi_{R0}=0$
and
$\psi_{R0} + i\eta_2 \psi_{L0}=0$. 
With $\eta_1= -\eta_2$, these two conditions 
are not compatible with each other.) 

This distinction between $\mathscr{P}^2=1$ and $\mathscr{P}^2=-1$ are directly related 
to the $\mathbb{Z}_2$ classification in class D+R$_+$ and 
to the absence of SPT phases in class D+R$_-$. 
For class D+R$_-$ ($P^2=-1$), the fermion zero modes are not allowed,
and hence the corresponding cross-cap state is constructed entirely 
in terms of fermionic oscillator modes (non-zero modes).
It can be checked easily that this cross-cap state is anomaly free. 
On the other hand, 
for class D+R$_+$ ($P^2=+1$), 
the construction of the corresponding cross-cap state is more non-trivial
because of the presence of fermion zero modes. 
We will see the action of the fermion number parity operator on the cross-cap 
state, $|C (\eta_1, \eta_2)\rangle$ with $\eta_1=\eta_2$, 
give rise to an anomalous phase (sign), indicative of the expected $\mathbb{Z}_2$
classification.

\paragraph{Construction of cross-cap states}
Let us now construct the cross-cap state focusing on $\eta_1 = \eta_2= \eta$:
\begin{align}
  \left[
   \psi^a_L (\sigma_2)- i \eta \psi^a_R (\sigma_2+\beta)
  \right]
  |C, \eta\rangle &=
  0,
\nonumber \\
 \left[
  \psi^a_R (\sigma_2)+ i \eta \psi^a_L (\sigma_2+\beta)
 \right]
 |C, \eta\rangle &=0,
\end{align}
where $a=1,\ldots, N_f$.
When $\eta_1=\eta_2$, the cross-cap condition is compatible with the periodic boundary
condition for the fermion field in $\sigma_1$ direction. 
Within the zero mode sector, the cross-cap condition reads 
\begin{align}
  \left[
   \psi^a_{0L} - i \eta \psi^a_{0R} 
  \right]
  |C, \eta\rangle 
  &=
  0, 
\end{align}
where $\psi^a_{0L/R}$ is the zero mode for the $a$-th flavor,
satisfying $(\psi^{a}_{0L/R})^2=1$.

In the following, we construct the cross-cap states within the zero-mode sector explicitly,
together with a reference state. 
[Recall the important role played by the reference state for the case of the bosonic SPT phases
discussed below Eqs.\ (\ref{csymmetry:momentum}) and (\ref{1})].
The choice of the reference state, however, is far from obvious, 
due to the degeneracy in the zero mode sector. 
We will illustrate this point by contrasting two constructions. 

In the first construction, 
we construct the Hilbert space of the zero modes
by considering
the following fermion creation/annihilation operators:  
\begin{align}
f^{\dag}_{a} &= 
\frac{1}{{2}}(\psi^a_{0L} + {i}\psi^a_{0R}), 
\quad
f^{\ }_{a} =
\frac{1}{{2}}(\psi^a_{0L} - {i}\psi^a_{0R}),
\end{align}
and the Fock vacuum $|0_f\rangle$ 
of the $f$-fermions. 
The cross-cap state $|C, \eta=+\rangle$ is then nothing
but 
$|0_f\rangle$ itself, 
\begin{align}
 |C, \eta =+\rangle =  e^{ i\phi_+}|0_f\rangle.  
\end{align}
On the other hand,
the cross-cap state $|C, \eta=-\rangle$ can be constructed as
\begin{align}
 |C, \eta =-\rangle = e^{i\phi_-} \prod_{a=1}^{N_f} f^{\dag}_a  |0_f\rangle. 
\end{align}
[N.B.
In the above representation of $|C,\eta\rangle$,
the ambiguous phases $\phi_{\pm}$ are not fixed by the cross-cap condition.
These phases will not affect our later analysis, and hence will be set to zero henceforth. 
One common convention for the phase is 
$
|C, \eta = + \rangle = e^{-i\frac{\pi}{8}}|0_f\rangle 
$, 
$
|C, \eta = - \rangle = e^{i\frac{\pi}{8}} f^{\dag}|0_f \rangle, 
$
where we focus on the case of single flavor $N_f=1$ for simplicity.
Then we notice the following identities  
%\begin{align}
$
\psi_{0L} |C, \eta \rangle = e^{-i\eta \frac{\pi}{4}}|C, -\eta \rangle 
$, 
$
\psi_{0R} |C, \eta \rangle = e^{-i\eta \frac{\pi}{4}}|C, -\eta \rangle.
$
%\label{phase choice}
%\end{align}
One motivation for this phase convention is that 
it well compares with the operator product expansions of the Ising CFT, 
\cite{CFTbook} 
$\psi_{L} \sigma \sim e^{i\frac{\pi}{4}}\mu$,
$\psi_{L} \mu \sim  e^{-i \frac{\pi}{4}}\sigma$,
$\psi_{R} \sigma \sim e^{-i\frac{\pi}{4}} \mu$, 
$\psi_{R} \mu \sim e^{i \frac{\pi}{4}} \sigma$,
where $\sigma$ and $\mu$ are the Ising spin operator and the disorder operator, respectively. 
]

Alternatively, 
when $N_f=\mbox{even}$, 
one can introduce the following fermion creation operators 
(see, for example, Ref. \onlinecite{Bergman1997}):
\begin{align}
d^{\dag}_{L j} = 
\frac{1}{{2}}(\psi^{2j-1}_{0L} + {i} \psi^{2j}_{0L}), 
\quad
{d}^{\dag}_{R j} =
\frac{1}{{2}}({\psi}^{2j-1}_{0R} + {i} {\psi}^{2j}_{0R}), 
\end{align}
and the Fock vacuum $|0_d\rangle$ annihilated by $d_{Lj}$ and $d_{Rj}$.  
We observe that
$
d_{Lj}- {i}{d}_{Rj}
=
f_{2j-1} - {i} f_{2j} 
$,
$(d^{\dag}_{jL} - i d^{\dag}_{jR}) |0_f\rangle=0$, 
and
\begin{align}
 |C, + \rangle = |0_f\rangle = \prod_j (d^{\dag}_{jL} - i d^{\dag}_{jR}) |0_d\rangle. 
\end{align}
Similarly,
\begin{align}
|C, - \rangle = \prod_{a=1}^{N_f} f^{\dag}_a  |0_f\rangle = \prod_j (d^{\dag}_{j R} - i d^{\dag}_{j L}) |0_d\rangle.
\label{2} 
\end{align} 
(Similar construction is possible even for $N_f=\mbox{odd}$
by adding an extra Majorana fermion as in Ref. \onlinecite{Fidkowski2011}.
We will not dwell on this point
as the identification of a quantum anomaly when $N_f=\mbox{odd}$ 
is rather straight forward and does not depend on the choice of the reference state.)
One important feature of the second construction is the clear factorization of
the vacuum $|0_d\rangle$ into the left- and right-moving sectors, 
$|0_d\rangle = |0_d\rangle_L\otimes |0_d\rangle_R$,
as it is annihilated by $d_{Lj}$ and $d_{Rj}$ separately.
The entire Hilbert space built out of $|0_d\rangle$ also factorizes into the left- and right-moving sectors.
This factorization allows us to introduce the action of reflection (time-reversal) in a transparent way. 
Such factorization is also expected for general construction of 
cross-cap states and boundary states in boundary conformal field theories. 
(See, for example, Ref. \onlinecite{Bates2006}.)

\subsection{Fermion number parity - $\mathbb{Z}_2$ classification in class D+R$_{+}$}

We now ask the properties of the cross-cap states 
under the action of symmetry generators. 
As a start, let us consider the fermion number parity.
In the tree channel, its explicit form within the zero mode sector is given by
\begin{align}
 \mathscr{G}_f 
 =
 (i \psi^1_{0L}\psi^1_{0R})
 (i \psi^2_{0L}\psi^2_{0R})
 \cdots
 (i \psi^{N_f}_{L}\psi^{N_f}_{0R}). 
\end{align}
The fermion number parity acting on the cross-cap states gives
\begin{align}
\mathscr{G}_f|C, \pm \rangle = (\pm)^{N_f} |C, \pm\rangle. 
\label{sign D+R+}
\end{align}
Therefore, 
when $N_f=\mbox{even}$, there is no anomaly.
On the other hand,
when $N_f=\mbox{odd}$,
one would conclude,
$|C, -\rangle$ is anomalous
while 
$|C, +\rangle$ is not.
This is consistent with the $\mathbb{Z}_2$ classification of 
(2+1)d topological superconductors. 
in symmetry class D+R$_{+}$. 

Upon closer inspection, however, Eq.\ (\ref{sign D+R+}) would look strange 
since the two cross-cap states $|C, \pm \rangle$
should be treated on the equal footing: 
If the cross-cap $|C, \pm\rangle$ is obtained 
by twisting $\mathscr{P}$, 
$|C, \mp\rangle$ is obtained from simply from $\mathscr{G}_f \mathscr{P}$. 
Hence, both cross-cap states $|C, \pm\rangle$ belong/correspond to the same phase. 
In fact, it should be noted that there is a phase ambiguity
in defining the cross-cap states and the fermion number parity operator. 
In the above analysis, we implicitly made a particular choice where  
the fermion number parity of the ground state $|0_f\rangle$ is $+1$. 
In principle, one could assign a different fermion number parity eigenvalue,
e.g., by modifying the definition of the fermion number parity operator,
$\mathscr{G}_f\to -\mathscr{G}_f$.
Alternatively, instead of using $f^{\dag}, f$,
one could define $c:=f^{\dag}$ and $c^{\dag}:=f$, 
which leads to 
$|C, -\rangle =|0_c\rangle$
and 
$|C,+\rangle =\prod_a c^{\dag}|0_c\rangle$. 
In this convention, it is tempting to 
claim 
$\mathscr{G}_f|C,-\rangle = +|C,-\rangle$
while 
$\mathscr{G}_f|C,+\rangle = (-1)^{N_f}|C,+\rangle$. 
Thus, there is some ambiguity when deducing
the fermion number parity eigenvalue.
We have encountered similar ambiguity when 
dealing with CP symmetric bosonic SPT phases. 
Such ambiguity of the fermion number parity eigenvalue of the ground state,
however, does not affect our conclusion,
since, independent of the phase choice,  
when $N_f=\mbox{odd}$, 
we cannot make both $|C,\pm\rangle$ anomaly-free.
Hence we conclude the $\mathbb{Z}_2$ classification of symmetry class D+R$_{+}$.

\subsection{Time-reversal - $\mathbb{Z}_8$ classification in class BDI + R$_{++}$}

We now include time-reversal symmetry. 
After $\pi/2$ rotation of spacetime, 
a time-reversal operator is transformed into a parity operator (unitary) (denoted by $\widetilde{\mathscr{P}}$ in the following).
Correspondingly to two time-reversal operators $\mathscr{T}^2=\pm 1$ in the loop channel picture, 
there are two kinds of parity operators in the tree channel picture.
They act on the fermion zero modes as:
\begin{align}
\widetilde{\mathscr{P}} \psi_{0L} \widetilde{\mathscr{P}}^{-1} = \psi_{0R},
 \quad
\widetilde{\mathscr{P}} \psi_{0R} \widetilde{\mathscr{P}}^{-1} =  \psi_{0L}, 
\end{align}
and
\begin{align}
\widetilde{\mathscr{P}} \psi_{0L} \widetilde{\mathscr{P}}^{-1} = \psi_{0R},
 \quad
 \widetilde{\mathscr{P}} \psi_{0R} \widetilde{\mathscr{P}}^{-1} =  -\psi_{0L}. 
\end{align}
Explicitly, they can be written as
\begin{align}
\widetilde{\mathscr{P}}
 = 
 e^{i\delta}
 \prod_{a}
 \frac{1}{\sqrt{2}} \left(\psi_{0L} + \psi_{0R} \right)^a 
\end{align}
and
\begin{align}
\tilde{ \mathscr{P}} =
 e^{i\delta}
 \prod_a 
 \frac{1}{\sqrt{2}}
 \left(
 1-
 \psi_{0L}\psi_{0R}
 \right)^a,  
\end{align}
respectively,
where $e^{i\delta}$ is an unknown phase factor and will be discussed in more detail shortly. 
The first parity operator does not preserve the cross-cap condition with $\eta_1=\eta_2$,
so we will focus on the second one. 
One also verifies
\begin{align}
\widetilde{\mathscr{P}}^2 = e^{2i\delta} (i)^{N_f} \mathscr{G}_f,
\label{projective}
\end{align}
and 
$ \mathscr{G}_f \widetilde{\mathscr{P}}= \widetilde{\mathscr{P}}\mathscr{G}_f$.
%The phase $(i)^{N_f}$ can be removed by redefining $\mathscr{G}_f\to -\mathscr{G}_f$ when $N_f=2\times \mbox{odd}$.)

Let us now calculate the action of $\widetilde{\mathscr{P}}$ on
the cross-cap states. 
By using the representation in terms of the $f$-fermions, 
$\psi_{0L}\psi_{0R} = i (2 f^{\dag} f - 1)$, 
$\widetilde{\mathscr{P}}$ can be written as
\begin{align}
\widetilde{\mathscr{P}} =
e^{i\delta} 
 \prod_a 
 \frac{1}{\sqrt{2}}
 \left[
 1- i (2 n_a -1)
\right],  
\end{align}
where $n_a = f^{\dag}_a f_a$. 
Then, the action of $\widetilde{\mathscr{P}}$ on the cross-cap states
is given by 
\begin{align}
\widetilde{\mathscr{P}} |C, +\rangle
&=
%\widetilde{\mathscr{P}} |0\rangle_f =
e^{i\delta}
 \prod_a 
 \frac{1}{\sqrt{2}}
 \left[
 1- i (2 n_a -1)
\right]  
|0_f\rangle
\nonumber \\%%%%%
&=
e^{i\delta}
 \prod_a 
 \frac{1}{\sqrt{2}}
 \left[
 1+ i 
\right]  
|0_f\rangle
=
e^{ +i\frac{\pi}{4}N_f}
e^{i\delta}
|C, +\rangle, 
\nonumber \\
\widetilde{\mathscr{P}} |C, -\rangle
&=
e^{i\delta}
 \prod_a 
 \frac{1}{\sqrt{2}}
 \left[
 1- i 
\right]  
|C, -\rangle
=
e^{ -i\frac{\pi}{4}N_f} 
e^{i\delta}
|C, -\rangle. 
\end{align}

The relative phase between 
$\widetilde{\mathscr{P}}|C, +\rangle$
and 
$\widetilde{\mathscr{P}}|C, -\rangle$
is $e^{ + i \pi N_f/2}$, 
which 
is independent of the choice of $e^{i\delta}$ (the choice of the action of $\widetilde{\mathscr{P}}$ on the reference state),
and
vanishes when $N_f = 4 \times \mbox{integer}$. 
In other words, 
one cannot make both cross-cap states anomaly-free unless $N_f = 4\times \mbox{integer}$. 
One then immediately concludes the classification is {\it at least} $\mathbb{Z}_4$.

On the other hand,
as we have seen for the case of the bosonic SPT phases, 
a proper choice of the phase $e^{i\delta}$, if exists, leads to a refined classification. 
If we choose $|0_f\rangle$ as the reference state and demand $|0_f\rangle$ transform trivially under $\widetilde{\mathscr{P}}$, 
$\widetilde{\mathscr{P}}|0_f\rangle = |0_f\rangle$, 
we obtain $\mathbb{Z}_4$ classification. 
%(This choice, in fact, is the same choice as above.)
Alternatively,  
if we choose $|0_d\rangle$ as the reference state, 
and demand $\widetilde{\mathscr{P}}|0_d\rangle=|0_d\rangle$,
which may be obtained from
$\widetilde{\mathscr{P}}|0_d\rangle_{L,R} = |0_d\rangle_{R,L}$,
\begin{align}
 \widetilde{\mathscr{P}}|C, + \rangle
 &= \widetilde{\mathscr{P}} |0_f\rangle \nonumber \\
 &= \widetilde{\mathscr{P}} \prod_{j=1}^{N_f/2} (d^{\dag}_{jL} - i d^{\dag}_{jR} )|0_d\rangle
 \nonumber \\
 &=
 \prod_{j=1}^{N_f/2} (d^{\dag}_{jR} +i d^{\dag}_{jL} )|0_d\rangle
\nonumber \\ 
 &=
 (i)^{N_f/2}|0_f\rangle. 
\end{align}
I.e., $e^{i\delta} = 1$.
It can also be checked, straightforwardly,  
\beq
\widetilde{\mathscr{P}}|C, - \rangle =  (i)^{N_f/2} |C, - \rangle, 
\eeq  
following Eq.\ \eqref{2}. 
Thus, with this choice, the two cross-cap states $|C,\eta\rangle$ can be both made anomaly free
only when $N_f=8\times \mbox{integer}$, i.e., $\mathbb{Z}_8$ classification. 

As a final comment, we provide a yet another point of view by using Eq.\ (\ref{projective}). 
Equation (\ref{projective}) suggests, within the zero mode sector, the symmetry is realized projectively. 
The ``unwanted'' phase $e^{2i \delta} (i)^{N_f}$ can be removed by  
choosing $e^{i\delta} = e^{- i \pi N_f/4}$.
However, with this choice, the reference state now acquires an anomalous phase 
$\widetilde{\mathscr{P}}|0_d\rangle = e^{ -i \pi N_f/4} |0_d\rangle$. 
This conflict between the two demands,
one to represent the symmetry group non-projectively 
and 
the other to make the reference state transform trivially under $\widetilde{\mathscr{P}}$, 
can be considered as a form of quantum anomaly.

%In fact, one does not need an explicit form of $\widetilde{\mathscr{P}}$:
%\begin{align}
%(\psi_L - i \psi_R )|+\rangle =0,
%\quad
%|-\rangle =(\psi_L + i \psi_R )|+\rangle  
%\end{align}
%Action of $R$ on $|-\rangle$ is
%\begin{align}
% \widetilde{\mathscr{P}} |+\rangle &= e^{iB}|+\rangle
% \nonumber \\%%%%%
% \widetilde{\mathscr{P}}|-\rangle
% &=
%(-i)(\psi_L - i \psi_R )\widetilde{\mathscr{P}} |+\rangle  
%=
%(-i) e^{iB} |-\rangle
%\end{align}
%The relative phase factor
%$(-i)$ is indicative of $\mathbb{Z}_4$ classification. 

\section{Conclusion}

There are three known ways of symmetry breaking in nature. 
First, a symmetry can be broken {\it explicitly} by adding a symmetry-breaking perturbation to the Hamiltonian.
Second, a symmetry can be broken {\it spontaneously} in many-body systems and in field theories 
because of interactions. 
These two forms of symmetry breaking occur both in classical and quantum systems. 
The third way of symmetry breaking is more subtle and happens only in quantum many-body systems;
A symmetry can be {\it anomalous}, meaning it can be broken because of quantum effects.
In discussing SPT phases, we are to distinguish different quantum phases of matter, all respecting the same set of symmetries. 
Therefore, by definition, the Landau paradigm, while powerful in discussing phases with spontaneous symmetry breaking,
cannot be applied to SPT phases. 
On the other hand, as we demonstrated, 
anomalous symmetry breaking (quantum anomalies) 
can be useful in diagnosing and perhaps even classifying SPT phases.  

The idea of using quantum anomalies to characterize topological phases of matter 
goes back to Laughlin's thought experiment (Laughlin's gauge argument) in the quantum Hall states. 
Our methodology can be considered as a proper generalization of Laughlin's gauge argument to SPT phases
protected by parity (reflection) and other symmetries. 
In Laughlin's thought experiment, 
quantized charge pumping caused by flux threading (large gauge transformation) 
characterizes the quantum Hall effect even in the presence of interactions and disorder. 
Within edge theories of the quantum Hall effect, the charge pumping by a large gauge transformation 
appears as a quantum anomaly, i.e., non-conservation of the electric $U(1)$ charge. 

The edge theories of SPT phases (if exist) differ from the edge theories of the quantum Hall systems.
First, quite generically, edge theories supported by SPT phases are non-chiral (having vanishing thermal Hall conductance).
Second, edge theories of SPT phases may be protected by discrete symmetries, and may not have a continuous $U(1)$ symmetry. 
For these reasons, diagnosing edge theories of SPT phases requires to identify
quantum anomalies (breaking down of symmetries of SPT phases by quantum effects) of a subtler kind than in the quantum Hall effect.   
By twisting parity in edge theories of (2+1)d topological phases,
we obtain a cross-cap state and investigate possible quantum anomalies in terms of the cross-cap state. 
This method is applied explicitly to the several examples, including bosonic and fermionic SPT phases, 
and reproduces the known classifications in Refs.\ \onlinecite{Hsieh_CPT, yao2013interaction}. 
In conclusion, 
we have provided a way to gauge the spatial symmetries to efficiently diagnose SPT phases, 
which is beyond the cohomology classification tables.  
We close with a few comments.

(i)
In our approach, it is crucial to fix the action of symmetry operators on a reference state in the tree-channel picture. 
In particular, possible phase ambiguity should be fixed, although 
in many cases a partial result is obtained even without fixing this ambiguity. 
While we have provided a proposal to fix this phase ambiguity by specifying reference states,  
it is desirable to have a more convincing and convenient method. 
One possible approach to this problem is to use the state-operator correspondence  
and assign the phase which is consistent with the operator algebra of the edge theories.
For example, such argument is used in Ref. \onlinecite{Gimon1996} in somewhat similar context. 
%[See also comments near Eq.\ (\ref{phase choice}).] 
Following this type of approach is left for the future research.

(ii)
We treated SPT phases where parity and time-reversal are in conflict quantum mechanically, 
in that enforcing parity by orientifolding leads to breakdown of time-reversal,
e.g., class BDI+R$_{++}$ topological superconductors. 
We choose to twist parity since twisting unitary symmetry appears to be more straightforward,
while, in principle, we could equally consider twisting by time-reversal. 
In fact, after exchanging the role of space and time coordinates and when considering 
%the action of time-reversal on 
cross-cap states, 
what appears to be time-reversal symmetry in the original, loop-channel picture
is represented as a parity (or $CP$) symmetry in the tree-channel picture. 
This indicates that by exchanging the role of space and time coordinates,
twisting time-reversal is neither more nor less than twisting parity, and should lead to an orientifold theory.
Our analysis in this paper thus provides an insight into SPT phases protected, not by parity, but by time-reversal symmetry 
and other symmetry, 
e.g., (2+1)d time-reversal symmetric topological insulators (the quantum spin Hall effect), 
which are protected by $U(1)_V$ and time-reversal symmetries.
Orientifolds of gapless edge states discussed in this work  
may provide a complementary view to, e.g.,  
Ref. \onlinecite{Chen_Gauging_Tsymmetry},
which discusses a way to gauge time-reversal in gapped bulk SPT phases 
by using the tensor network representation of quantum states. 

(iii) 
In our approach, 
the use of cross-cap states, upon going  from the loop to tree channel picture,  
most clearly demonstrates how quantum anomalies may appear.
In doing so, we exchange the role of space and time coordinates.
The fictitious time-evolution and the fictitious Hilbert space in the tree channel picture
is in fact akin to 
the row-to-row transfer matrix and to the auxiliary space of matrix product states (MPSs), respectively.
The projective representation of the symmetry group of SPT phases on the auxiliary space of MPSs has been successfully used 
to diagnose and classify (1+1)d gapped SPT phases. 
Our use of cross-cap states and the associated anomalous phases have some similarities with the 
classification of (1+1)d gapped SPT phases by MPSs. 

(iv)
In our previous calculations \cite{Hsieh2014}, 
the full partition functions on the Klein bottle are computed 
explicitly and checked for the presence/absence of quantum anomalies.
In the current reformulation in the tree channel,
there is no need to compute the partition function,
although it is possible to compute the partition function by using cross-cap states.
In addition, we do not rely on the continuous $U(1)$ symmetry, and hence can treat 
a wider class of SPT phases that are not discussed in Ref. \onlinecite{Hsieh2014}.  

The tree-channel formulation also somewhat liberates us from the unwanted reliance on relativistic and conformal symmetries
when discussing topological classification of phases of matter. 
As an MPS can be constructed for an arbitrary (1+1)d (gapped as well as gapless) quantum states,
cross-cap states can in principle be constructed for general edge theories which lack relativistic invariance.
If it comes to compute the full partition function, the lack of the relativistic invariance may be inconvenient,
as the Hamiltonian in the loop and tree channels do not agree. 
The partition function in such situations would not be expressed in terms of well-known functions such as the Jacobi theta functions. 

The importance, necessity and limitations of relativistic and conformal invariance, 
when following our approach to SPT phases,  
are however not entirely understood currently, and are left for the future studies. 
It would be interesting to note that, in constructing cross-cap states and, similarly, 
in constructing conformal invariant boundary states in boundary conformal field theories, 
the Virasoro algebra plays a major role. 

%Before finishing up the paper, one may ask how to detect numerically the anomalous phases of the crosscap states under the symmetry actions, which have been discussed in our paper. As the edge theory of the SPT phase alone cannot be regularized properly, we need to introduce the bulk in the crosscap geometry. It is not difficult to imagine to introduce another spatial direction $x_3$ in the spacetime Fig.1 along which we attach the bulk of the SPT states, e.g., we attach the bulk of the SPT phase along $x_3 >0$. After introducing the bulk, we have an edge state on the crosscap geometry at $x_3 = 0$. Now we can apply the symmetries and look at the anomalous phases generated by the symmetry actions. Though it looks feasible to do the numerics, we leave the concrete realization of numerical studies to confirm our result in future.

\acknowledgements
We thank Xie Chen and Andreas Ludwig 
for helpful discussion. 
%The authors acknowledge support from the NSF grants DMR-1064319 (G.Y.C). 
%GYC thanks Andreas Ludwig and Xie Chen for useful discussions. 
This work is supported 
by 
the NSF under Grant No. DMR-1064319 (GYC), No. DMR 1408713 (GYC), and DMR-1455296 (SR), 
the Brain Korea 21 PLUS Project of Korea Government (GYC),
and Alfred P. Sloan foundation (SR).

\appendix

\section{Bosonic SPT phases}
\label{Bosonic SPT phases app}

In this appendix, we will complete the classifications of (2+1)d bosonic SPT phases
protected by  
$P\times TC, CP\times T, T\times C$ and $T\times C\times P$.
The methodology for the classification is already explained in the main text. 
In this Appendix, to simplify our analysis, we will make one more assumption:
the CPT-theorem. 
If we limit our focus to topological classification of relativistic systems, 
the CPT-theorem allows us to relate 
various SPT phases protected by different sets of symmetries.
For example, with the CPT-theorem, 
one can convert, e.g., the classification problem of
bosonic SPT phases with $T$-symmetry into the classification problem of 
bosonic SPT phases with $CP$-symmetry. 
While the restriction to relativistic systems may sound too stringent, 
in Ref.\ \onlinecite{Hsieh_CPT}, it was shown that 
classifications of free fermionic SPT phases by K-theory or the Clifford algebra 
do satisfy the CPT-equivalence in general. I.e., 
different fermionic SPT phases whose symmetries are CPT-conjugate to each other 
are classified by the same topological charge. 
While we do not have such systematic understanding for the case of bosonic 
SPT phases, which are necessarily interacting, 
within microscopic analysis of gapping potentials performed in Ref. \onlinecite{Hsieh_CPT},
such CPT relations among different SPT phases still exist for bosonic SPT phases. 
Our analysis presented below, which makes use of cross-cap states and assumes 
the CPT-theorem, also confirm these results, although it should be emphasized 
that the assumption of the CPT-theorem can be avoided if one wishes.

\paragraph{$P\times TC$ symmetry}

We illustrate the use of CPT-theorem
by considering the edge theory of bosonic SPT phases protected by $P \times TC$.
As $P$-symmetry, we consider the case with $m_p=0$
(and the corresponding cross-cap state $|C_p (n_p,0)\rangle$ in Eq.\ \eqref{xcapstate:p}). 
On the other hand,  
$TC$-symmetry, which is a ''spatial`` symmetry in time and anti-unitary, is given by 
\begin{align}
\mathscr{TC}
&: \phi \rightarrow - \phi + n_{TC} \pi R,
\quad
\theta \rightarrow \theta + \frac{m_{TC} \pi \alpha'}{R} \nonumber\\ 
&:(x_1,x_2) \rightarrow (x_1, -x_2), 
\end{align}
where $n_{TC}, m_{TC} \in \{0, 1\}$. 

(i) As the first step,
we convert $TC$-symmetry into its CPT partner, 
i.e., a parity symmetry $P'$: 
\begin{align}
\mathscr{P}'&: \phi \rightarrow  \phi + n_{TC} \pi R, 
\quad
\theta \rightarrow -\theta + \frac{m_{TC} \pi \alpha'}{R} \nonumber\\ 
&:(x_1,x_2) \rightarrow (-x_1, x_2), 
\end{align}
which is unitary. 
Thus we have the bosonic theory with the two parity symmetries $P \times P'$ after this mapping.

(ii) Next, 
we form a non-spatial unitary symmetry by combining $P$ and $P'$, $X = P \cdot P'$: 
\begin{align}
\mathscr{X}&: \phi \rightarrow \phi + (n_{p} + n_{TC}) \pi R,
\quad
\theta \rightarrow \theta + \frac{m_{TC} \pi \alpha'}{R} \nonumber\\
&:(x_1,x_2) \rightarrow (x_1,x_2). 
\end{align}
Because $X$ is non-spatial in $(x_1,x_2)$-space, 
it should be non-spatial in $(\sigma_{1}, \sigma_{2})$-space as well (see Fig.\ \ref{fig:xCap}). 
Furthermore, it is straightforward to check that $X$ leaves the cross-cap conditions 
by $P$ and $P'$ \eqref{xcapcondition:p} invariant .  

%Thus, the problem of diagnosing the edge theory with $P \times P'$ can be reduced  
%into two separate problems,  in which we have $P \times X$ and $P' \times X$. 

(iii) As the third step, 
we twist $P$ and $P'$ to construct the cross-cap states and study if they are invariant under $X$ or not. 
By twisting $P'$-symmetry, we form the cross-cap state 
$|C_{p'} (n_{TC}, m_{TC}) \rangle$. 
If either $|C_p (n_{p},0) \rangle$ or $|C_{p'} (n_{TC}, m_{TC}) \rangle$ 
is not invariant under $X$, 
then the corresponding theory with $P \times TC$ is anomalous:
\begin{align}
\mathscr{X} |C_p (n_{p},0) \rangle &= (-1)^{n_{p}m_{TC}}  |C_p (n_{p},0) \rangle,
\nonumber\\
\mathscr{X} |C_{p'} (n_{TC}, m_{TC}) \rangle &= (-1)^{n_{TC} m_{TC}} |C_{p'} (n_{TC}, m_{TC}) \rangle. 
\end{align}
Thus the edge theory is anomalous if $n_{p}m_{TC}=1$ {\it or} $n_{TC}m_{TC} =1$. 
To conclude, we have the following triplet $(n_{p}, n_{TC}, m_{TC})$ for the anomalous states. 
\begin{align}
(n_{p}, n_{TC}, m_{TC}) & = (1,0,1), (0,1,1), (1,1,1).  
\end{align}
With these procedure, 
we conclude that the theory with the symmetry $P \times TC$ 
is anomalous if its CPT partner, the state with $P \times P'$ symmetry, is anomalous. 
The result agrees with Ref. \onlinecite{Hsieh_CPT}, 
where all the possible perturbations which can potentially gap out the edge theory
are enumerated.

\paragraph{$CP \times T$ symmetry}
The cross-cap state for the SPT phase with $CP \times T$ is 
$|C_{cp} (n_{cp},m_{cp}) \rangle$ (Eq.\ \eqref{xcapstate:cp}),  
where $(n_{cp}, m_{cp}) \in \{ 0, 1\}$. 
%Now we consider $T$-symmetry which is non-on-site in {\it time} and is {\it anti-unitary}. 
%\begin{align}
%T&: \phi \rightarrow  \phi  \nonumber\\ 
%&: \theta \rightarrow -\theta + \frac{m_{T} \pi \alpha'}{R} \nonumber\\ 
%&:(x_1,x_2) \rightarrow (x_1, -x_2) 
%\end{align}
%Here $m_{T} \in \{0, 1\}$. 
We find the CPT partner $CP'$ of $T$-symmetry
\begin{align}
\mathscr{CP}'&: \phi \rightarrow  -\phi,
\quad
\theta \rightarrow \theta + \frac{m_{T} \pi \alpha'}{R} \nonumber\\ 
&:(x_1,x_2) \rightarrow (-x_1, x_2), 
\end{align}
which is unitary. 
We again form a non-spatial unitary symmetry by combining $CP$ and $CP'$. 
We call the symmetry as $Z = CP \cdot CP'$: 
\begin{align}
\mathscr{Z}&: \phi \rightarrow \phi + n_{cp} \pi R,
\quad
\theta \rightarrow \theta + \frac{(m_{T}+m_{cp}) \pi \alpha'}{R} \nonumber\\
&:(x_1,x_2) \rightarrow (x_1,x_2) 
\end{align}
$Z$ is non-spatial 
in $(\sigma_{1}, \sigma_{2})$-space (see Fig.\ \ref{fig:xCap}). We can also show that $Y$ leaves the cross-cap conditions by twisting $CP$ and twisting $CP'$ \eqref{xcapcondition:cp} invariant.  
Using the $CP'$-symmetry, we form another cross-cap state 
$|C_{cp'} (0, m_{T}) \rangle$. 
If $|C_{cp'} (0,m_{T}) \rangle$ or 
$|C_{cp} (n_{cp}, m_{cp}) \rangle$ is not invariant under $Z$, 
the corresponding theory with $CP \times T$ is anomalous. 
\begin{align}
\mathscr{Z} |C_{cp'} (0,m_{T}) \rangle &= (-1)^{n_{cp}m_{T}}  |C_{cp'} (0,m_{T}) \rangle,
\nonumber\\
\mathscr{Z} |C_{cp} (n_{cp}, m_{cp}) \rangle &= (-1)^{n_{cp} m_{cp}} |C_{cp} (n_{cp}, m_{cp}) \rangle.
\end{align}
Thus the edge theory is anomalous if $n_{cp}m_{T}=1$ {\it or} $n_{cp}m_{cp} =1$. 
We conclude the following triplet $(m_{T}, n_{cp}, m_{cp})$ for the anomalous states. 
\begin{align}
(m_{T}, n_{cp}, m_{cp}) & = (1,1,0),
(0,1,1),
(1,1,1). 
\end{align}
This result agrees with Ref. \onlinecite{Hsieh_CPT}. 

\paragraph{$P\times T$}
This case is analyzed in Sec.\ \ref{Bosonic SPT phases} without
using CPT theorem.
The CPT partner of $T$-symmetry,
$CP'$, is given by 
\begin{align}
\mathscr{CP}'&: \phi \rightarrow  -\phi
\quad
\theta \rightarrow \theta + \frac{m_{T} \pi \alpha'}{R} \nonumber\\ 
&:(x_1,x_2) \rightarrow (-x_1, x_2).
\end{align}
By combining $P$ and $CP'$,
we form a non-spatial unitary symmetry,
$Y \equiv P \cdot CP'$, 
which acts on the bosonic fields as 
\begin{align}
\mathscr{Y}&: \phi \rightarrow -\phi + n_{p} \pi R,
\quad \theta \rightarrow -\theta + \frac{m_{T} \pi \alpha'}{R} \nonumber\\
&:(x_1,x_2) \rightarrow (x_1,x_2).
\end{align}
$Y$ is non-spatial and unitary in $(\sigma_{1}, \sigma_{2})$-space (Fig.\ \ref{fig:xCap}). 
It can also be checked that $Y$ leaves the cross-cap conditions
by $P$ \eqref{xcapcondition:p} and $CP'$ \eqref{xcapcondition:cp} 
invariant.  
Twisting the $CP'$-symmetry, 
we form another cross-cap state 
$|C_{cp'} (0, m_{T}) \rangle$. 
The action of $Y$ on
the cross-cap states is given by
\begin{align}
\mathscr{Y} |C_p (n_{p},0) \rangle &= (-1)^{n_{p}m_{T}}  |C_p (n_{p},0) \rangle,
\nonumber\\
\mathscr{Y} |C_{cp'} (0, m_{T}) \rangle &= (-1)^{n_{p} m_{T}} |C_{cp'} (0, m_{T}) \rangle. 
\end{align}
Thus the edge theory is anomalous if $n_{p}m_{T}=1$, i.e., $(n_{p}, m_{T}) = (1,1)$.
This agrees with Ref.\ \onlinecite{Hsieh_CPT}. 

\paragraph{$T \times C$ symmetry}

Here in this case, there is no apparent $P$-symmetry.
We can however map $T$-symmetry into its CPT partner, 
i.e., $CP$-symmetry. 
Then we can twist $CP$-symmetry to obtain a cross-cap state and act $C$ on the cross-cap state to diagnose the stability of the theory. 
%We start with $T$-symmetry. The symmetry is non-on-site in time and anti-unitary. 
%\begin{align}
%T&: \phi \rightarrow  \phi + n_{T}\pi R \nonumber\\ 
%&: \theta \rightarrow -\theta + \frac{m_{T} \pi \alpha'}{R} \nonumber\\ 
%&:(x_1,x_2) \rightarrow (x_1, -x_2) 
%\end{align} 
The CPT partner of $T$-symmetry, $CP$-symmetry, is defined as  
\begin{align}
\mathscr{CP}&: \phi \rightarrow  -\phi + n_{T}\pi R,
\quad
\theta \rightarrow \theta + \frac{m_{T} \pi \alpha'}{R} \nonumber\\ 
&:(x_1,x_2) \rightarrow (-x_1, x_2) 
\end{align} 
By twisting this $CP$-symmetry, we obtain the cross-cap state 
$|C_{cp} (n_{T}, m_{T}) \rangle$ \eqref{xcapstate:cp}. 
On the other hand, $C$-symmetry \eqref{csymmetry:phase} is identified with $n_{c}=m_{c}=0$. 
Under the action of $C$-symmetry, 
the cross-cap condition \eqref{xcapcondition:cp} is invariant,
while we find,  by using \eqref{csymmetry:momentum}, 
\beq
\mathscr{C} |C_{cp} (n_{T}, m_{T}) \rangle = (-1)^{n_{T}m_{T}}| C_{cp} (n_{T},m_{T}) \rangle.
\eeq
Thus the theory with the parity symmetry with $n_{T}=1, m_{T}=1$ and $C$ is anomalous.
\cite{Hsieh_CPT} 

\paragraph{$T \times C \times P$ symmetry}

%For the symmetry group $Z_2^{\mathrm{T}}\times Z_2^{\mathrm{C}}\times Z_2^{\mathrm{P}}$, we have the following (gauge-inequivalent) transformations:
%\begin{align}
%T&: \phi \rightarrow  \phi + n_{T}\pi R  \nonumber\\ 
%&: \theta \rightarrow -\theta + \frac{m_{T} \pi \alpha'}{R} \nonumber\\ 
%&: (x_1,x_2) \rightarrow (x_1, -x_2) 
%\end{align}
%\begin{align}
%C:&\phi \rightarrow - \phi  \nonumber\\
%:&\theta \rightarrow -\theta  \nonumber\\ 
%:& (x_1,x_2) \rightarrow (x_1,x_2)
%\end{align}
%\begin{align}
%P:&\phi \rightarrow \phi + n_{p}\pi R \nonumber\\
%:&\theta \rightarrow -\theta + \frac{m_{p}\pi \alpha'}{R}\nonumber\\
%:&(x_{1},x_{2}) \rightarrow (\ell -x_{1},x_{2})
%\end{align}
%
The symmetry group $T \times  C \times P$ 
is CPT-equivalent to 
$CP' \times  C \times P $ 
or 
$P' \times C \times P$, 
where 
\begin{align}
\mathscr{CP}'&: \phi \rightarrow  -\phi + n_{T}\pi R,
\quad
\theta \rightarrow \theta + \frac{m_{T} \pi \alpha'}{R} \nonumber\\ 
&:(x_1,x_2) \rightarrow (-x_1, x_2),
\nonumber \\
\mathscr{P}'&: \phi \rightarrow  \phi + n_{T}\pi R,
\quad
\theta \rightarrow -\theta + \frac{m_{T} \pi \alpha'}{R} \nonumber\\ 
&:(x_1,x_2) \rightarrow (-x_1, x_2).
\end{align}
We can choose a set $\{P, P', CP'\}$ 
(other choices can be $\{P, P', CP\}$, $\{P, CP, CP'\}$, etc.) 
that generates the symmetry group $P' \times C \times P$,  
and consider their corresponding cross-cap states
$|C_{p} (n_{p}, m_{p}) \rangle$, $|C_{p'} (n_{T}, m_{T}) \rangle$ and $|C_{cp'} (n_{T}, m_{T}) \rangle$. 
To study the SPT phases (and their group structure) 
with the group $P' \times C \times P $, 
we can check how these cross-cap states transform under the non-spatial symmetries $g_i$
which are generated by combining symmetries from $\{P, P', CP'\}$, 
such as $X\equiv P\cdot P'$, $Y\equiv P\cdot CP'$, and $C$.
We can choose $\{X, Y\}$,  $\{X, C\}$, or  $\{Y, C\}$ as a "complete" set of non-spatial symmetries. 
Here we consider $\{Y, C\}$, and their action on the cross-cap states are given by:
\begin{align}
\mathscr{Y} |C_p (n_{p}, m_{p}) \rangle &= (-1)^{n_{p}m_{T}}  |C_p (n_{p}, m_{p}) \rangle,
\nonumber\\
\mathscr{C} |C_p (n_{p}, m_{p}) \rangle &= (-1)^{n_{p}m_{p}} |C_p (n_{p}, m_{p}) \rangle,
\nonumber\\
\mathscr{Y} |C_{p'} (n_{T}, m_{T}) \rangle &= (-1)^{n_{T}m_{p}}  |C_{p'} (n_{T}, m_{T}) \rangle,
\nonumber\\
\mathscr{C} |C_{p'} (n_{T}, m_{T}) \rangle &= (-1)^{n_{T}m_{T}} |C_{p'} (n_{T}, m_{T}) \rangle,
\nonumber\\
\mathscr{Y} |C_{cp'} (n_{T}, m_{T}) \rangle &= (-1)^{n_{p} m_{T}} |C_{cp'} (n_{T}, m_{T}) \rangle,
\nonumber\\
\mathscr{C} |C_{cp'} (n_{T}, m_{T}) \rangle &= (-1)^{n_{T}m_{T}} |C_{cp'} (n_{T}, m_{T}) \rangle.
\label{sym_on_Xcap_states_TCP_2}
\end{align}
The system is in a nontrivial SPT phase if there exists at least one symmetry-noninvariant cross-cap state. 
From \eqref{sym_on_Xcap_states_TCP_2}, this occurs when 
\begin{align}
[n_{\mathrm{T}}, m_{\mathrm{T}}, n_{\mathrm{P}}, m_{\mathrm{P}}]= 
&[0, 0, 1, 1],\ [0, 1, 1, 0],\ [0, 1, 1, 1], \nonumber\\
&[1, 0, 0, 1],\ [1, 1, 0, 0],\ [1, 1, 0, 1], \nonumber\\
&[1, 0, 1, 1],\ [1, 1, 1, 0],\ [1, 1, 1, 1].
\label{boson_T_C_P_nontrivialSPT_1_Appendix}
\end{align}
Putting two identical copies of each phase above together 
results a trivial phase, i.e., $[n_{\mathrm{T}}, m_{\mathrm{T}}, n_{\mathrm{P}}, m_{\mathrm{P}}]^2=[\text{trivial}]$. 
Here, when we put two phases together, the resulting phase, say,
$[n_{\mathrm{T}}^1, m_{\mathrm{T}}^1, n_{\mathrm{P}}^1, m_{\mathrm{P}}^1]
\oplus
[n_{\mathrm{T}}^2, m_{\mathrm{T}}^2, n_{\mathrm{P}}^2, m_{\mathrm{P}}^2]$, 
has the corresponding cross-cap states 
$|C_{\Omega_i} (n_{\Omega_i}, m_{\Omega_i})\rangle=
|C_{\Omega_i}^1 (n_{\Omega_i}^1, m_{\Omega_i}^1)\rangle
\otimes|C_{\Omega_i}^2 (n_{\Omega_i}^2, m_{\Omega_i}^2)\rangle$, 
which are the direct (tensor) product of the cross-cap states of the original two phases.
On the other hand, all phases above are inequivalent [we can not obtain a trivial phase by putting any two different elements in the set (\ref{boson_T_C_P_nontrivialSPT_1_Appendix}) together].
However, phases shown in (\ref{boson_T_C_P_nontrivialSPT_1_Appendix}) are not all possible nontrivial SPT phases with symmetries $T \times C \times P$; putting different phases above together might result other nontrivial SPT phases that are not listed above. To determine the group structure of these phases, we observe that putting three phases in any vertical or  horizontal line of the array (\ref{boson_T_C_P_nontrivialSPT_1_Appendix}) together will result a trivial phase, or equivalently, we have
\begin{align}
&\oplusto{[0, 0, 1, 1]}{} \oplus \oplusto{[0, 1, 1, 0]}{} = \oplusto{[0, 1, 1, 1]}{} \nonumber\\
&\equalto{[1, 0, 0, 1]}{} \oplus \equalto{[1, 1, 0, 0]}{} = \equalto{[1, 1, 0, 1]}{} \nonumber\\
&[1, 0, 1, 1] \oplus [1, 1, 1, 0] = [1, 1, 1, 1].
\label{boson_T_C_P_nontrivialSPT_2_Appendix}
\end{align}
From this, we see that all nontrivial phases can be generated by a specific set of four phases, 
say, $\{[0, 0, 1, 1],\ [0, 1, 1, 0],\ [1, 0, 0, 1],\ [1, 1, 0, 0]\}$ 
(there are $3\times 3=9$ equivalent choices for these four group generators); 
there are in total fifteen nontrivial SPT phases, which form a $\mathbb{Z}_2^4$ group.
\cite{Hsieh_CPT}

\bibliography{ref}

%\tableofcontents 

\end{document}